\documentclass[sigconf,nonacm]{acmart}

\usepackage{placeins}
\usepackage{xspace}
\usepackage{amsmath}
\usepackage{xurl}
\usepackage{bbding}
\usepackage{listings}
\hypersetup{hidelinks}
\usepackage[capitalize]{cleveref}
\usepackage{enumitem}
\usepackage{siunitx}

\usepackage{tabularx}
\usepackage{makecell}
\usepackage{ragged2e}

\usepackage{multirow}

\newcolumntype{L}[1]{>{\hsize=#1\hsize\linewidth=\hsize\RaggedRight\arraybackslash}X}
\newcolumntype{C}[1]{>{\hsize=#1\hsize\linewidth=\hsize\Centering\arraybackslash}X}
\newcolumntype{R}[1]{>{\hsize=#1\hsize\linewidth=\hsize\RaggedLeft\arraybackslash}X}
\newcommand{\papertablestyle}{%
  \fontsize{8}{10}\selectfont
  \setlength{\tabcolsep}{3.5pt}%
  \renewcommand{\arraystretch}{1.15}%
  \hyphenpenalty=10000\relax
}

\newcommand{\KVC}{KV Cache\xspace}

\newcommand{\RoofModel}{Transformer Roofline Model\xspace}

\newcommand{\projectname}{TokenSimV3}
\newcommand{\project}{\projectname\xspace}
\setcopyright{none}

\begin{document}

\title[\projectname]{\projectname: Full-Stack Simulation for Agentic LLM Serving and Hardware–Software Co-Design}
\makeatletter
\newcommand{\paper@authorrows}{}
\newcommand{\paper@affiliationlist}{}

\newcommand{\declarePaperAffiliation}[4]{%
  \@namedef{paper@affiliation@#1}{%
    \affiliation{\institution{#2}\city{#3}\country{#4}}%
  }%
  \g@addto@macro\paper@affiliationlist{\textsuperscript{#1}\,#2, #3, #4\par}%
}

\newcommand{\paper@authorcell}[3]{%
  \begin{minipage}[t]{.245\textwidth}
    \centering
    {\@authorfont #1\textsuperscript{#3}\par}%
    {\small\nolinkurl{#2}\par}%
  \end{minipage}%
}

\newcommand{\paperAuthor}[3]{%
  \author{#1}%
  \@for\paper@id:=#3\do{\@nameuse{paper@affiliation@\paper@id}}%
  \email{#2}%
  \g@addto@macro\paper@authorrows{\paper@authorcell{#1}{#2}{#3}}%
}

\declarePaperAffiliation{1}{Peking University}{Beijing}{China}
\declarePaperAffiliation{2}{Beijing Tongming Lake Information Technology Application Innovation Center}{Beijing}{China}
\declarePaperAffiliation{3}{Independent~Researcher}{Beijing}{China}

\paperAuthor{Zhuohang Bian}{zhbian26@stu.pku.edu.cn}{1,2}
\paperAuthor{Songhan Huang}{2400012702@stu.pku.edu.cn}{1}
\paperAuthor{Zhuoran Li}{zhuoranli@stu.pku.edu.cn}{1}
\paperAuthor{Feiyang Wu}{fywu25@stu.pku.edu.cn}{1}
\g@addto@macro\paper@authorrows{\par\smallskip}
\paperAuthor{Chengkai Sheng}{2500011438@stu.pku.edu.cn}{1}
\paperAuthor{Teng Ma}{stmatengss@gmail.com}{3}
\paperAuthor{Yun Liang}{ericlyun@pku.edu.cn}{1}
\paperAuthor{Youwei Zhuo}{youwei@pku.edu.cn}{1,2}

\if@ACM@nonacm
  \if@ACM@anonymous\else
    \renewcommand{\@mkauthors}{%
      \global\setbox\mktitle@bx=\vbox{%
        \hsize=\textwidth
        \unvbox\mktitle@bx
        \medskip
        \centering\paper@authorrows\par
        \medskip
        \normalsize\paper@affiliationlist
        \bigskip
      }%
    }%
  \fi
\fi
\makeatother

\begin{abstract}
Evaluating large language model (LLM) serving policies and emerging hardware architectures directly on physical multi-node clusters is slow and prohibitively expensive.
Simulation provides a fast alternative, yet existing simulators largely target single-node execution with flat memory and independent requests, offering little help in real deployments governed by distributed communication, multi-tier \KVC{} management, and stateful agentic workflows.
We present \project, a fast, extensible full-stack simulator for multi-node agentic and disaggregated LLM serving.
First, \project reproduces the distributed agent serving environment across heterogeneous attention models and a three-tier \KVC{} hierarchy spanning GPU memory, host DRAM, and shared cluster storage.
Second, to avoid exhaustive profiling, \project unifies sparse local measurements with physical scaling laws to extrapolate across million-token contexts and multi-node collectives.
Validated on production GB200/GB300 clusters, \project predicts multi-node agent throughput within $7.1\%$--$9.9\%$ error and single-node time-to-first-token within $0.20\%$--$7.47\%$ error up to $1\text{M}$ context, while running up to $57\times$ faster than existing serving simulators and $553\times$ faster than packet-level network simulators.
We further use \project to uncover system design insights across programmable \KVC{} policies, Prefill--Decode and Attention--FFN disaggregation, Groq LPX accelerators, and High-Bandwidth Flash (HBF) storage.
\project is publicly available at \url{https://github.com/pku-lemonade/TokenSim}.
\end{abstract}
\maketitle
\hypersetup{pdfsubject={Preprint},pdfpublisher={}}

\section{Introduction}
\label{sec:intro}

\begin{figure}[H]
\centering
\includegraphics[width=\columnwidth]{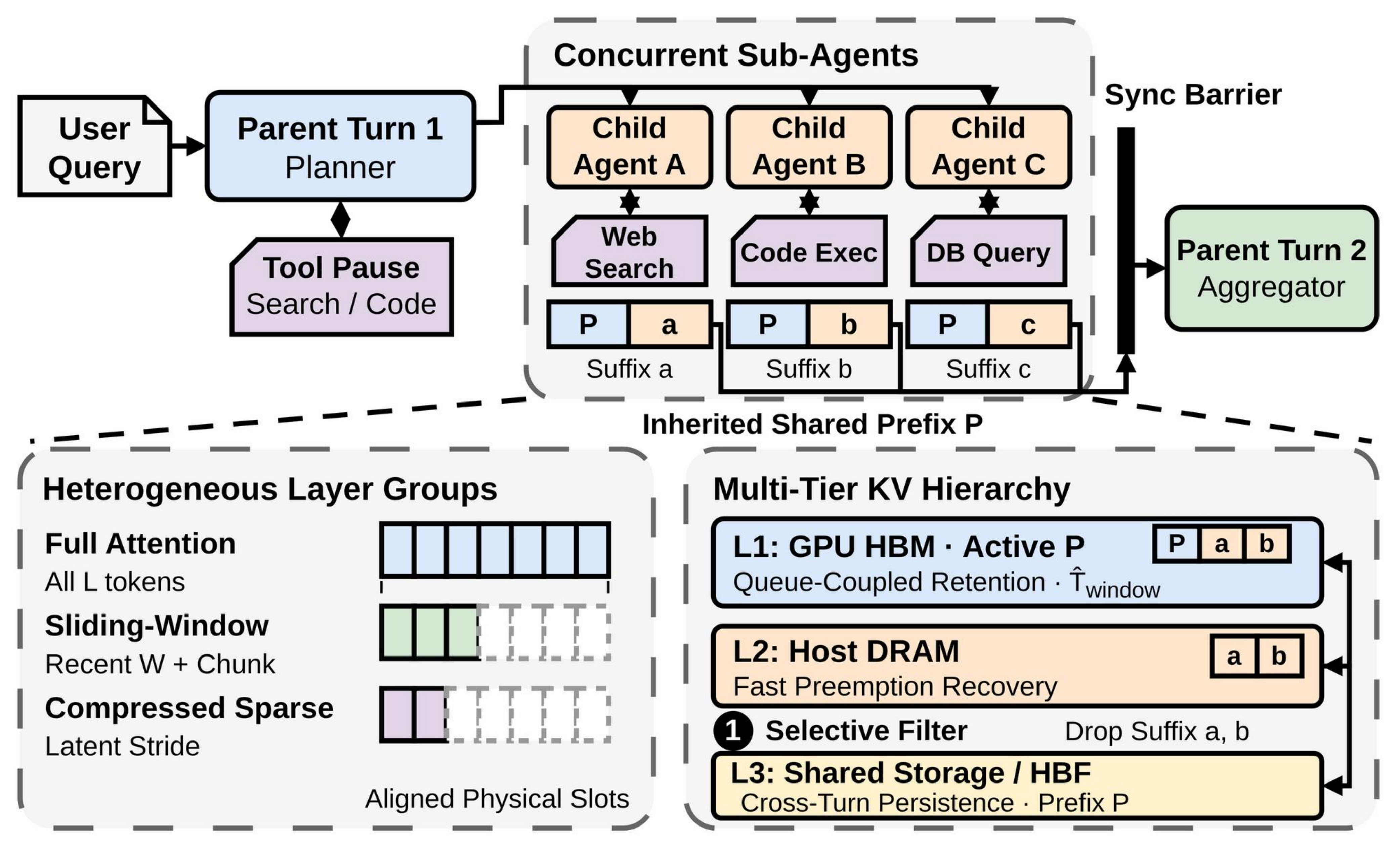}
\vspace{-10pt}
\caption{
Stateful multi-turn agent workflow coupled to heterogeneous attention layer groups and a three-tier \KVC{} hierarchy.
}
\label{fig:agent_workflow}
\vspace{-10pt}
\end{figure}

Production large language model (LLM) serving has evolved into stateful, multi-turn agentic workflows running massive Mixture-of-Experts (MoE) and long-context models across multi-node clusters~\cite{vllm,sglang}.
As illustrated in Figure~\ref{fig:agent_workflow}, parent--child sub-agent branches and tool pauses accumulate million-token contexts across heterogeneous full, sliding-window, and compressed attention layers~\cite{deepseekv4,gemma4}, driving frequent \KVC{} movement across L1 GPU HBM, L2 host DRAM, and L3 shared cluster storage~\cite{cachedattention,mooncake}.
Serving these workloads at scale requires tuning \textbf{multi-node agent serving stacks} across hybrid parallelisms (TP~\cite{megatron}, DP~\cite{orca}, EP~\cite{deepseekv3}, PP~\cite{gpipe}), speculative decoding~\cite{dflash,dspark}, Prefill--Decode (P/D) disaggregation~\cite{distserve,splitwise}, Attention--FFN (AF) disaggregation~\cite{step3,megascale-infer}, and multi-tier caching.
Architects also seek to evaluate \textbf{emerging compute and storage architectures}, such as SRAM-centric Groq LPX accelerators~\cite{groq-tsp,groq3_lpx_architecture} and High-Bandwidth Flash (HBF)~\cite{hbf}, under realistic agent workloads.
Sweeping these coupled software configurations on physical clusters is prohibitively costly, and emerging hardware rarely provides accessible multi-node testbeds, so architects rely on fast, accurate simulation prior to deployment.

\begin{figure*}[t]
\centering
\begin{minipage}[c]{0.49\textwidth}
\centering
\includegraphics[width=\linewidth]{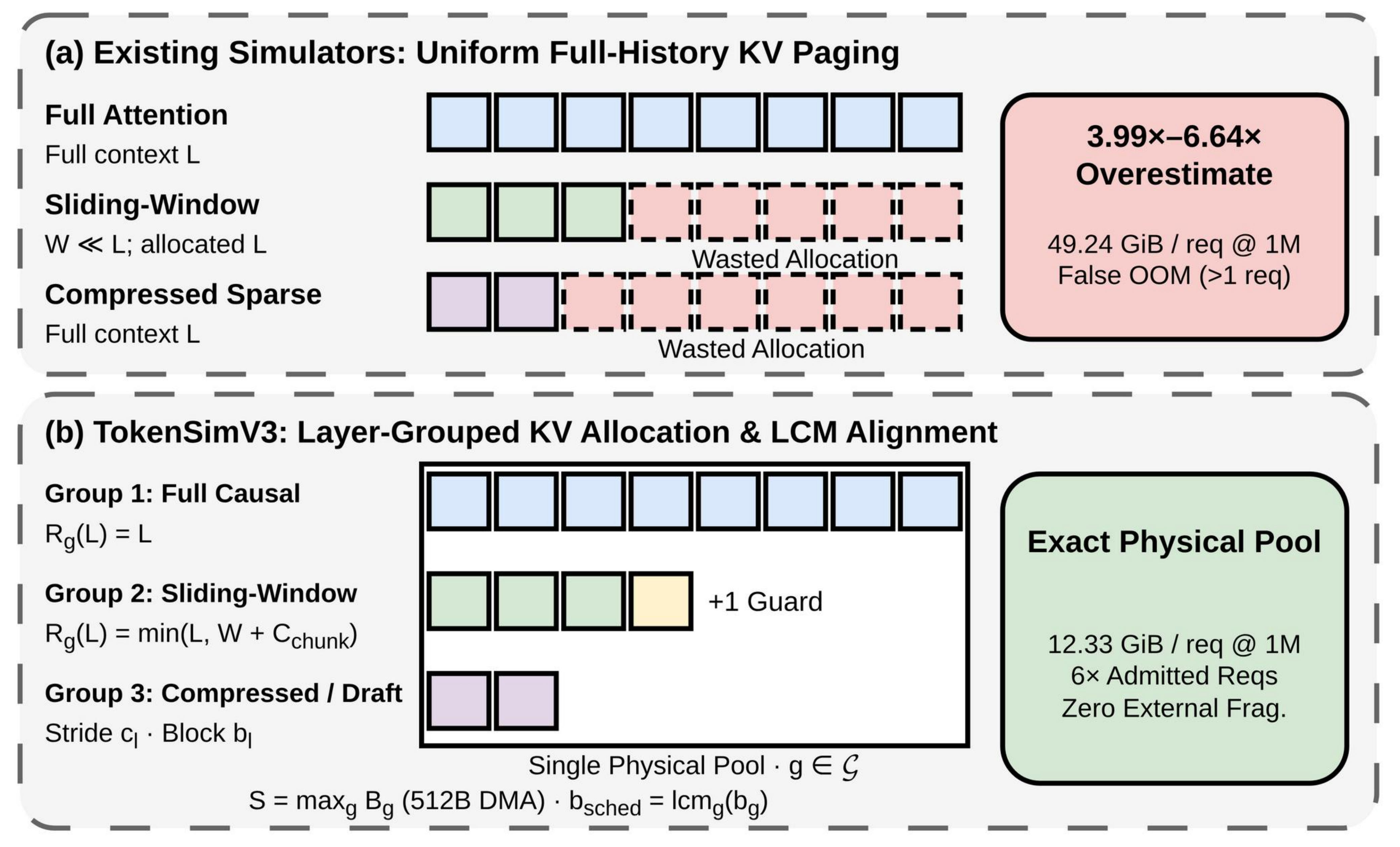}
\end{minipage}\hfill
\begin{minipage}[c]{0.49\textwidth}
\centering
\includegraphics[width=\linewidth]{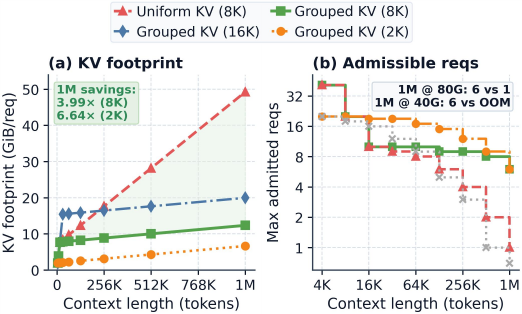}
\end{minipage}
\caption{
\textbf{Left:} Uniform full-history \KVC{} paging vs.\ \project's layer-grouped allocation.
\textbf{Right:} Per-request \KVC{} footprint (a) and admissible concurrent requests (b) on DeepSeek-V4-Pro up to $1\text{M}$ context.
}
\label{fig:kv_capacity}
\vspace{-10pt}
\end{figure*}

However, existing LLM serving simulators~\cite{vidur,llmservingsim2,agentservesim,aiconfigurator} remain limited by flat single-pool memory and independent requests, unscalable multi-node network models, and isolated operator estimators, exposing three fundamental challenges:

\noindent\textbf{Stateful agent workflows and heterogeneous, multi-tier \KVC{} lifecycles.}
Existing single-engine simulators~\cite{vidur,llmservingsim,llmservingsim2,agentservesim} that model isolated arrivals over a single GPU pool cannot capture how tool pauses and parent--child synchronization barriers drive \KVC{} demotions, evictions, and restores across the L1--L3 cache hierarchy (Figure~\ref{fig:agent_workflow}).
Modeling this multi-tier hierarchy also exposes heterogeneous per-layer \KVC{} lifecycles, where uniform full-context paging wastes physical slots on sliding-window and compressed attention layers~\cite{deepseekv4,gemma4} (Figure~\ref{fig:kv_capacity}, left).
For DeepSeek-V4-Pro up to $1\text{M}$ context (Figure~\ref{fig:kv_capacity}, right), this uniform full-history assumption overestimates per-request \KVC{} footprints by $3.99\times$--$6.64\times$ ($49.24\,\text{GiB}$ vs.\ $12.33\,\text{GiB}$ at $8\text{K}$ chunk, Figure~\ref{fig:kv_capacity}(a)) and falsely restricts an $80\,\text{GiB}$ pool to $1$ concurrent request instead of $6$ (Figure~\ref{fig:kv_capacity}(b)).
Accurately evaluating concurrency limits and cache policies therefore requires coupling stateful agent execution with layer-grouped memory budgeting and a multi-tier cache state machine.

\noindent\textbf{Simulation accuracy and speed across multi-node and disaggregated scale.}
Scaling serving across multi-node TP, PP, wide EP, and disaggregated P/D or AF pools exposes inter-node communication contention and cross-pool synchronization bottlenecks that existing simulators struggle to capture accurately.
Coarse static tables~\cite{simai,vidur} incur $23.8\%$--$27.2\%$ collective error as node counts grow by ignoring shared per-node NIC injection bottlenecks, while rigid pipeline models fail to capture high-concurrency, out-of-order cohort arrivals that cause $67.8\%$ head-of-line stalls under AF jitter.
Conversely, packet-level network simulators~\cite{astrasim} and cycle-level hardware simulators~\cite{llmservingsim2} do not scale in simulation speed, taking minutes ($553\times$ slower) to simulate even a few dozen collectives at thousand-rank scale.
Rapid cluster-scale exploration therefore requires a closed-form, topology-aware hierarchical collective model coupled with event-driven cohort scheduling to capture NIC contention and cross-pool queuing in milliseconds.

\noindent\textbf{Isolated operator latency estimation methods.}
Existing simulators estimate operator execution time using one of three isolated paradigms~\cite{charon}, each of which breaks down in long-context, multi-node serving.
\emph{Empirical profiling} is noisy and difficult at ultra-long contexts and requires prohibitive sampling budgets to cover two-dimensional cached-prefill grids $(L_q, L_{\mathrm{kv}})$, incurring up to $17.0\%$ error when approximated by zero-history diagonal tables (Figure~\ref{fig:estimator_flow}, right).
\emph{Analytical roofline models}~\cite{aiconfigurator} and \emph{linear table extrapolation} fail to isolate realized hardware efficiency and fixed kernel launch overheads across long contexts, causing $27.5\%$--$144.7\%$ decode extrapolation error and $104.2\%$ $1\text{M}$-prefill error.
\emph{Machine-learning predictors}~\cite{vidur} incur $7.70\,\text{ms}$ per-query latency ($200\times$ slower) while failing beyond their training sequence lengths.
Overcoming these limits requires unifying sparse single-node measurements with boundary-anchored analytical scaling and cached-prefix correction to preserve hardware efficiency and launch costs during long-context extrapolation.

\noindent\textbf{\project design and contributions.}
We present \project (Figure~\ref{fig:arch}), a full-stack discrete-event simulation framework for multi-node agentic and disaggregated LLM serving, validated against production GB200/GB300 clusters, $32\times\text{H20}$ deployments, multi-node cloud fabrics, and a 256-chip Groq LP30 system, with four main contributions:
\begin{itemize}[leftmargin=*, nosep]
\item \textbf{Stateful agent and multi-tier \KVC{} simulation:} We couple multi-turn agent dependencies with layer-grouped memory allocation (preventing $3.99\times$--$6.64\times$ memory overestimation at $1\text{M}$ context) and a programmable L1--L3 \KVC{} hierarchy.
\item \textbf{Physics-anchored hybrid latency engine:} We unify sparse local measurements ($25\%$ budget) with physical scaling laws, achieving $0.40\%$ prefill error at $1\text{M}$ context and $5.2\%$--$9.3\%$ multi-node collective error while running $200\times$ faster than Random Forest predictors and $553\times$ faster than packet-level network simulators.
\item \textbf{Cluster validation and disaggregated serving insights:} \project tracks multi-node GB200/GB300 agent throughput within $7.1\%$--$9.9\%$ error (up to $57\times$ faster than existing serving simulators) and uncovers key design insights for programmable \KVC{} policy co-design, P/D ratio provisioning, and jitter-resilient AF disaggregation.
\item \textbf{Emerging hardware exploration:} Through modular compute, memory, and storage abstractions, \project enables rapid evaluation of emerging architectures under realistic agent workloads, such as SRAM-centric Groq LPX accelerators ($4.66\%$ error) and High-Bandwidth Flash (HBF) ($53.6\%$ lower agent completion time).
\end{itemize}
\section{Background and Motivation}
\label{sec:background}

\begin{table*}[!t]
\centering
\caption{Comparison of \project with existing LLM serving and distributed network simulators.}
\label{tab:simulator_capabilities}
\fontsize{8}{9.5}\selectfont
\setlength{\tabcolsep}{4.5pt}
\renewcommand{\arraystretch}{1.05}
\begin{tabularx}{\textwidth}{@{}lL{0.92}L{1.02}L{1.06}L{1.00}@{}}
\toprule
Simulator & Workload Runtime & \KVC{} Hierarchy \& Layout & Operator Timing & Disaggregation \& Hardware Scope \\
\midrule
Vidur~\cite{vidur} & Single-turn requests & L1 GPU paging & Dense tables + RF regressors & Colocated TP/PP \\
LLMServingSim 2.0~\cite{llmservingsim2} & Single-turn requests & L1/L2/CXL prefix cache & Cycle-level NPUs \& network & P/D, AF, TP/PP/EP \\
AgentServeSim~\cite{agentservesim} & Multi-turn agent workflows & L1 GPU prefix pool & Empirical operator models & Single-pool GPU serving \\
Charon~\cite{charon} & Operator execution graphs & Static graph buffer sizing & Profile, analytical \& ML models & Graph partitioning \& overlap \\
SimAI~\cite{simai} & Distributed execution traces & Static buffer allocation & Static comm.\ tables \& formulas & Multi-node collectives \\
ASTRA-sim 2.0~\cite{astrasim} & Distributed execution graphs & Packet/buffer state events & Packet-level network simulation & Hierarchical collectives \\
AISimulate~\cite{aisimulate,aiconfigurator} & Agent replay; static search & Tiered cache & Tables + interpolation + roofline & P/D, static AF \\
rLLM~\cite{rllm} & vLLM plugin replay & vLLM PagedAttention & Hardware operator models & Accelerator evaluation \\
Uni3Sim~\cite{uni3sim} & Mocked vLLM/SGLang & Tiered memory (HBFrame) & ArchPrism \& ProtoBridge & Multi-node GPU/NPU \\
\midrule
\textbf{\project} & \textbf{Stateful multi-agent DAGs} & \textbf{Grouped KV; programmable L1--L3} & \textbf{Sparse anchors + physical scaling} & \textbf{TP/EP/PP/DP, P/D, AF, LPX \& HBF} \\
\bottomrule
\end{tabularx}
\end{table*}

\subsection{Stateful Agentic Workflows and Disaggregated Serving}
\label{subsec:bg_agentic}

Modern LLM serving workloads consist of multi-turn agentic workflows that alternate between LLM generation, external tool calls, and parallel child sub-agent branching and synchronization~\cite{vllm,sglang}.
In production DeepSeek-V4 AgentX traces on $8\times\text{B200}$ and $8\times\text{B300}$ clusters, $61.7\%$ and $57.9\%$ of completed requests exceed $64\text{K}$ input tokens, and within the $64\text{K}$--$128\text{K}$ bin on B200, measured TTFT p90 drops from $100.53\,\text{s}$ at turns $1$--$3$ to $7.82\,\text{s}$ at turns $16+$ as later turns hit cached prefixes~\cite{cachedattention,mooncake}.
Long-context models interleave full causal, sliding-window (SWA), block-compressed sparse, and multi-token prediction (MTP) draft layers alongside weights, communication buffers, workspaces, activations, and CUDA graphs~\cite{deepseekv4,gemma4}; as Figure~\ref{fig:kv_capacity} shows, uniform full-history \KVC{} models overestimate memory by $3.99\times$--$6.64\times$ at $1\text{M}$ context.

To eliminate prefill--decode interference and expand \KVC{} capacity, production clusters adopt Prefill--Decode (P/D) disaggregation~\cite{distserve,splitwise} and MoE Attention--FFN (AF) disaggregation~\cite{step3,megascale-infer}.
However, optimal P:D worker splits shift sharply with compute ($r_C$) and interconnect bandwidth ($r_B$) ratios, while offloading routed MoE weights onto stateless FFN ($F$) pools forces repeated HBM weight streaming for small token micro-batches and exposes multi-layer pipeline stalls under network and TP jitter.
Furthermore, emerging architectures such as SRAM-centric LPX accelerators ($500\,\text{MB}$ on-chip SRAM at $150\,\text{TB/s}$)~\cite{groq-tsp,realscale,groq3_lpx_architecture} and multi-terabyte High-Bandwidth Flash (HBF)~\cite{hbf} introduce strict on-chip SRAM residency bounds, fine-grained launch overheads, and asymmetric flash I/O contention.

\subsection{Limitations of Existing Simulators}
\label{subsec:bg_operator}

Table~\ref{tab:simulator_capabilities} compares \project against representative LLM serving and network simulators.
Request-level simulators such as Vidur~\cite{vidur} and LLMServingSim 2.0~\cite{llmservingsim2} assume independent arrivals and uniform GPU paging, omitting multi-turn agent dependencies and layer-grouped \KVC{} allocation.
While AgentServeSim~\cite{agentservesim} and AISimulate~\cite{aisimulate,aiconfigurator} replay agent workflows, their fixed runtimes cannot compose speculative decoding, global expert parallelism, dynamic AF micro-batch aggregation, and programmable L1--L3 cache policies.
For latency estimation, empirical ML regressors~\cite{vidur,charon} incur millisecond-scale query overhead and cannot extrapolate beyond profiled lengths, whereas uncalibrated roofline scaling amplifies fixed kernel launch overhead at long contexts and diagonal prefill tables ignore historical \KVC{} reads.
Static collective tables~\cite{simai} miss shared NIC injection contention, while packet-level simulators~\cite{astrasim} require over $146\,\text{s}$ for $64$ collectives at $1024$ ranks.
A complementary class of framework-coupled simulators, including rLLM~\cite{rllm} and Uni3Sim~\cite{uni3sim}, attaches compute, memory, and packet-level interconnect models to live plugins or mocked execution paths of production engines such as vLLM and SGLang.
While mocking existing engine stacks reproduces current scheduling and paging behavior, it locks exploration into mechanisms already implemented inside those codebases.
\section{\project Architecture and Methodology}
\label{sec:method}

\subsection{Unified Discrete-Event Simulation Architecture}
\label{subsec:method_overview}

\begin{figure}[t]
\centering
\includegraphics[width=\columnwidth]{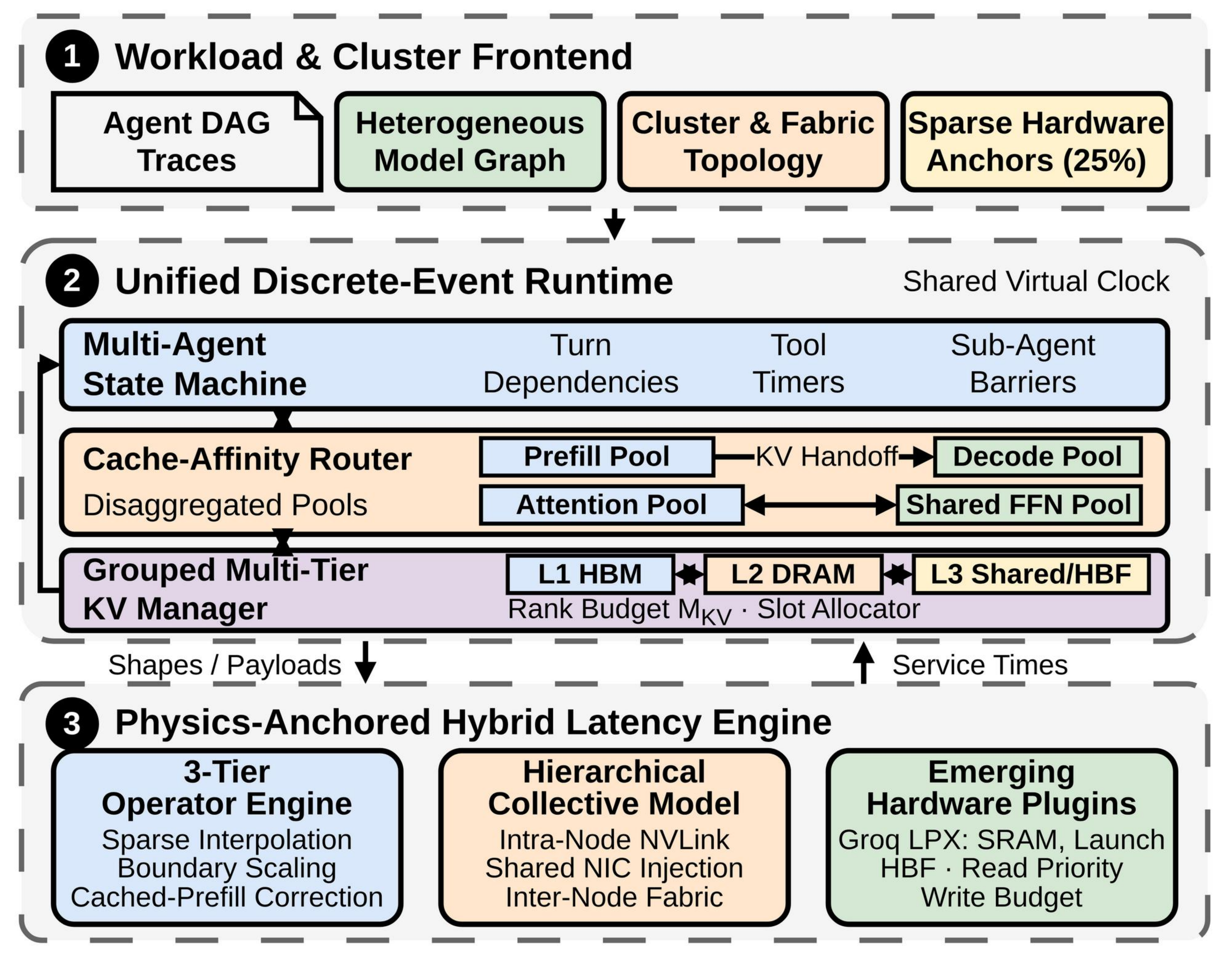}
\caption{
\project architecture: (1) workload and cluster frontend, (2) unified discrete-event runtime, and (3) physics-anchored hybrid latency engine.
}
\label{fig:arch}
\vspace{-10pt}
\end{figure}

\project uses discrete-event simulation~\cite{simpy} to couple agent workflows with disaggregated serving resources (Figure~\ref{fig:arch}).
As shown in Figure~\ref{fig:arch}(1), \project takes four modular inputs: (1) a model graph specifying per-layer attention mechanisms, MoE routing, and numerical precisions; (2) a cluster topology declaring compute ranks, memory tiers, and fabrics; (3) sparse operator and collective latency anchors; and (4) a stateful agentic trace with multi-turn dependencies, tool pauses, and prefix block hashes.
Within the discrete-event runtime (Figure~\ref{fig:arch}(2)) and hybrid latency engine (Figure~\ref{fig:arch}(3)), \project advances cluster state on a shared virtual clock as agent dependencies resolve, routers assign turns to pools, schedulers form prefill chunks or decode batches, interconnects transfer state, and operators execute.
The runtime separates intrinsic service time, computed by the hybrid latency engine, from queuing delay across GPU streams, PCIe DMA engines, NIC injection ports, and flash lanes, so changes to memory capacity, cache policies, or pool ratios dynamically reshape downstream arrivals and prefix hit rates.
Decoupling this discrete-event runtime from any fixed serving engine codebase lets researchers reproduce production batching and caching behavior while rapidly prototyping unimplemented cache policies, disaggregated pipelines, and emerging hardware without modifying vLLM or SGLang.

\subsection{Stateful Agent Replay and Layer-Grouped Memory Budgeting}
\label{subsec:method_kv}

\begin{figure}[t]
\centering
\includegraphics[width=\columnwidth]{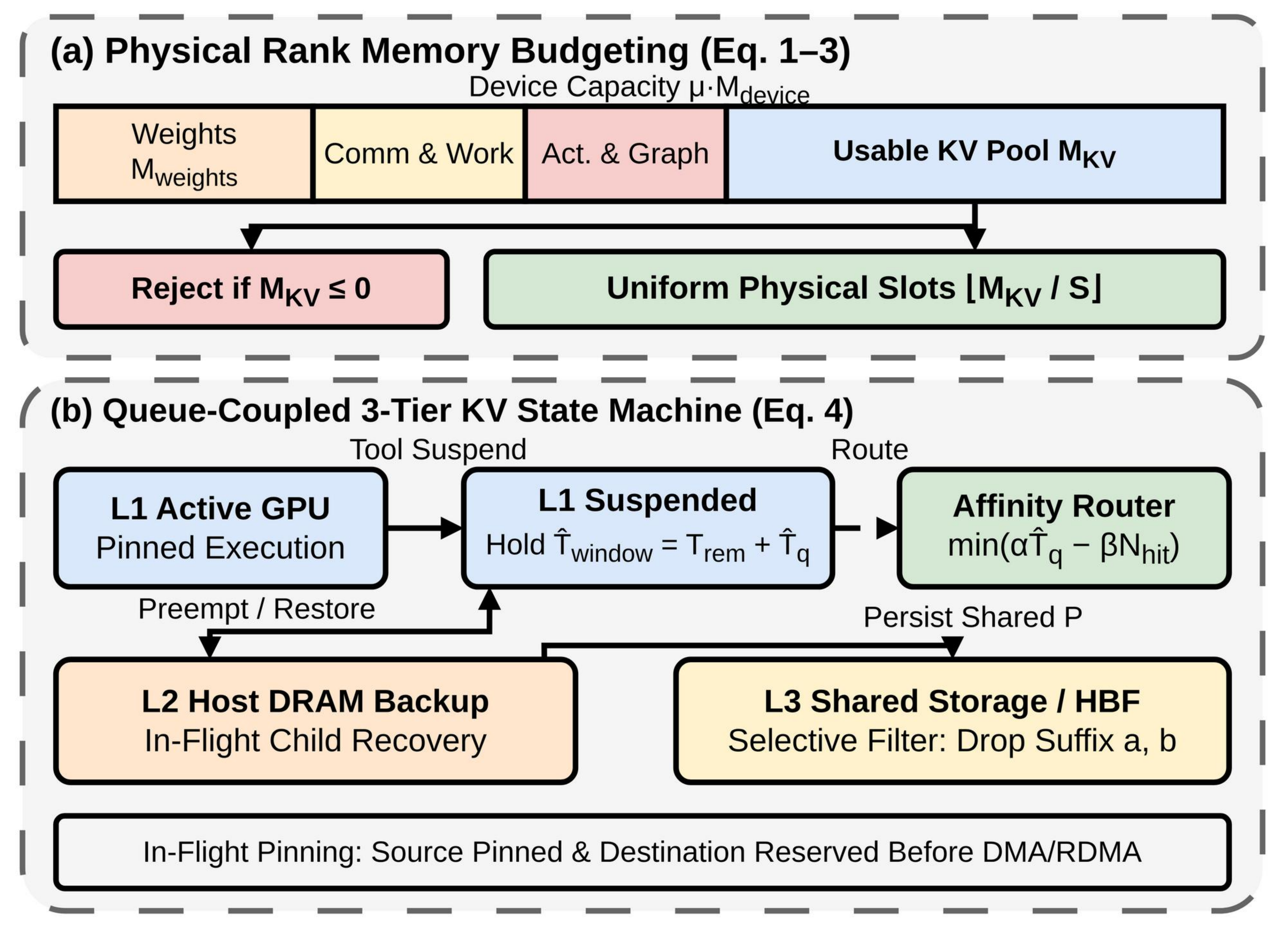}
\caption{
Runtime-coupled \KVC{} management: (a) rank memory budgeting and slot allocation; (b) three-tier lifecycle.
}
\label{fig:cache_lifecycle}
\vspace{-10pt}
\end{figure}

\noindent\textbf{Stateful multi-agent execution and rank memory budgeting.}
\project models each session as a directed acyclic graph (DAG) of dependent turns and synchronization barriers: branches alternate between on-cluster generation and off-cluster tool execution, while parent turns can spawn concurrent child branches sharing read-only parent prefixes and block at join barriers until child outputs return.
As shown in Figure~\ref{fig:cache_lifecycle}(a), before admitting requests, \project computes the usable \KVC{} byte budget $M_{\mathrm{KV}}$ on each rank by deducting role-specific weight shards $M_{\mathrm{weights}}$ (e.g., omitting routed MoE weights on Attention ranks in AF disaggregation), communication reserves, persistent workspaces, peak in-flight activations, and CUDA graphs from the scaled device capacity $\lfloor\mu M_{\mathrm{device}}\rfloor$, rejecting any placement with $M_{\mathrm{KV}} \le 0$.

\noindent\textbf{Heterogeneous layer-grouped \KVC{} allocation.}
To avoid the uniform full-history memory overestimation shown in Figure~\ref{fig:kv_capacity}, \project partitions layers into homogeneous cache groups $g \in \mathcal{G}$ sharing identical block width $b_l$, compression ratio $c_l$, retention window $w_l$, and byte width $s_l$ (Figure~\ref{fig:cache_lifecycle}(a)).
Within layer $l$, each page stores $b_l / c_l$ compressed entries rounded up to the $512$-byte DMA boundary ($P_l = 512\lceil (b_l/c_l)s_l / 512\rceil\,\mathrm{bytes}$).
Summing across layers in group $g$ yields group page bytes $B_g = \sum_{l \in g} P_l$, and a uniform physical slot stride $S = \max_{g \in \mathcal{G}} B_g$ manages all groups in a unified pool of $\lfloor M_{\mathrm{KV}} / S \rfloor$ slots without external fragmentation, aligned to scheduling granularity $b_{\mathrm{sched}} = \operatorname{lcm}_{g \in \mathcal{G}}(b_g)$.
At context length $L$ under prefill chunk budget $C_{\mathrm{chunk}}$, group $g$ retains $R_g(L) = L$ tokens if unwindowed ($w_g = 0$) and $R_g(L) = \min(L, w_g + C_{\mathrm{chunk}})$ if windowed ($w_g > 0$), requiring total slots:
\begin{equation}
N_{\mathrm{req}}(L)=\sum_{g \in \mathcal{G}}\left(\left\lceil\frac{R_g(L)}{b_g}\right\rceil+\mathbf{1}[w_g>0]\right),
\label{eq:hybrid_slots}
\end{equation}
where $\mathbf{1}[w_g > 0]$ reserves a rotating write-guard page for sliding-window updates.

\begin{figure*}[t]
\centering
\begin{minipage}[c]{0.49\textwidth}
\centering
\includegraphics[width=\linewidth]{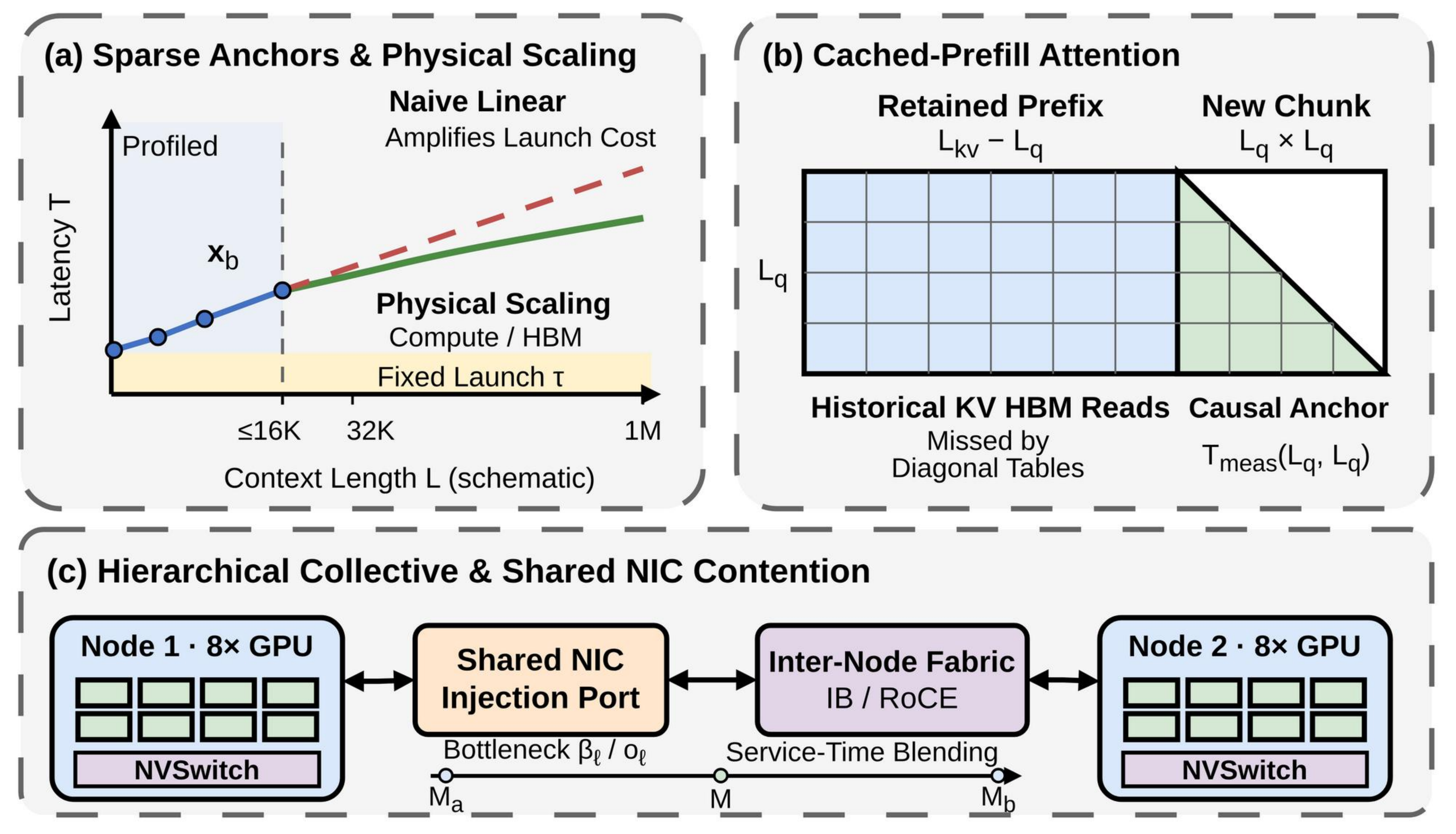}
\end{minipage}\hfill
\begin{minipage}[c]{0.49\textwidth}
\centering
\includegraphics[width=\linewidth]{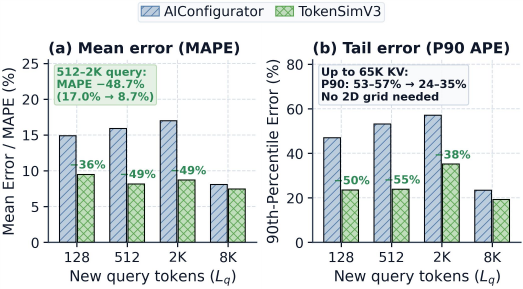}
\end{minipage}
\caption{
\textbf{Left:} Hybrid latency engine: (a) sparse interpolation and boundary-anchored scaling, (b) cached-prefill decomposition, and (c) hierarchical collective modeling.
\textbf{Right:} Off-diagonal cached-prefill prediction error: (a) mean error (MAPE) and (b) 90th-percentile tail error (P90 APE).
}
\label{fig:estimator_flow}
\vspace{-10pt}
\end{figure*}

\subsection{Programmable Multi-Tier \KVC{} Control}
\label{subsec:method_pkv}

Figure~\ref{fig:cache_lifecycle}(b) illustrates how \project tracks ordered prefix blocks across L1 device memory (HBM or LPX SRAM), L2 host DRAM over PCIe/NVLink-C2C, and L3 shared storage (distributed DRAM, CXL, or HBF/NVMe flash) via a Mooncake-style connector~\cite{mooncake}.
Each block transitions across five ownership states (\texttt{RESIDENT\_L1}, \texttt{PINNED\_IN\_FLIGHT}, \texttt{RESIDENT\_L2}, \texttt{RESIDENT\_L3}, and \texttt{EVICTED}), pinning in-flight sources and reserving destination slots during DMA and RDMA transfers.
To separate cache-policy exploration from inference-engine internals, \project exposes programmable policy hooks at agent turn boundaries (\texttt{on\_suspend}, \texttt{on\_resume}, \texttt{on\_evict}, and \texttt{on\_writeback}) that read application lifecycle hints, per-tree quota limits, and live worker pressure without modifying vLLM or SGLang.
These hooks compute a queue-coupled retention window $\widehat{T}_{\mathrm{window}}(r) = T_{\mathrm{tool\_rem}}(r) + \widehat{T}_{\mathrm{queue}}(w)$ from remaining tool wait and target prefill backlog to avoid premature L1$\to$L2 demotions.
They also decouple \emph{L2 preemption-recovery backups} (retained for active child branches so preemption avoids recomputation) from \emph{selective L2$\to$L3 persistence write-backs} (\texttt{selective\_write\_back}, filtering completed one-shot child suffixes at L2 eviction), while cache-affinity routing selects the worker $w^* = \arg\min_{w}(\alpha \widehat{T}_{\mathrm{queue}}(w) - \beta N_{\mathrm{hit}}(w))$ to balance queue delay against matched prefix tokens $N_{\mathrm{hit}}(w)$.

\subsection{Physics-Anchored Hybrid Operator and Collective Engine}
\label{subsec:method_latency}

\noindent\textbf{Hybrid operator timing and cached-prefill correction.}
To avoid costly, high-variance ultra-long-context profiling, \project extrapolates from sparse moderate-length profiles collected on one node.
As illustrated in Figure~\ref{fig:estimator_flow} (left, a), given operator shape vector $\mathbf{x}$ (batch size, query tokens $L_q$, retained context $L_{\mathrm{kv}}$, and hidden dimensions), \project resolves latency in three steps:
(1) returning exact matches or interpolating between sparse measurements $\mathcal{D}_{\mathrm{meas}}$ along each axis;
(2) extrapolating beyond boundary $\mathbf{x}_b$ by scaling the boundary anchor $T_{\mathrm{meas}}(\mathbf{x}_b)$ by calibrated analytical ratios~\cite{aiconfigurator}:
\begin{equation}
\widehat T(\mathbf{x})=T_{\mathrm{meas}}(\mathbf{x}_b)
\frac{A(\mathbf{x})}{A(\mathbf{x}_b)};
\label{eq:extrap}
\end{equation}
and (3) falling back to the calibrated \RoofModel when no local measurements exist ($\mathcal{D}_{\mathrm{meas}} = \emptyset$):
\begin{equation}
A(\mathbf{x})=a\left[\max\left(
\frac{F(\mathbf{x})}{P\eta_c},
\frac{D(\mathbf{x})}{B\eta_m}
\right)+\tau\right]+b,
\label{eq:roofline}
\end{equation}
where $F(\mathbf{x}), D(\mathbf{x})$ are FLOPs and bytes, $P, B$ are peak compute and bandwidth, $\eta_c, \eta_m$ are realized efficiencies, and $\tau, b$ capture fixed launch and tail overheads isolated inside $A(\mathbf{x}_b)$ during short-context calibration.
As shown in Figure~\ref{fig:estimator_flow} (left, b), to predict off-diagonal cached prefill ($L_{\mathrm{kv}} \gg L_q$) from zero-history diagonal anchors $T_{\mathrm{meas}}(L_q, L_q)$ without 2D profiling, \project scales $T_{\mathrm{meas}}(L_q,L_q)$ by $\max(1, A(L_q,L_{\mathrm{kv}})/A(L_q,L_q))$ via Equation~\ref{eq:extrap}, where $A(L_q, L_{\mathrm{kv}})$ sums causal attention within $L_q$ and historical \KVC{} reads over $L_{\mathrm{kv}} - L_q$.
Across $936$ off-diagonal cached-prefill shapes ($L_{\mathrm{kv}} > L_q$, up to $65\text{K}$), Figure~\ref{fig:estimator_flow} (right) shows that this decomposition reduces mean error (MAPE) from $17.0\%$ (AIConfigurator 1D lookup~\cite{aiconfigurator}) to $8.7\%$ and 90th-percentile tail error from $53\%$--$57\%$ to $24\%$--$35\%$ without profiling a 2D grid.

\noindent\textbf{Topology-aware hierarchical collective modeling.}
As shown in Figure~\ref{fig:estimator_flow} (left, c), for multi-node collectives across intra-node NVLink/NVSwitch and shared per-node NIC injection ports into inter-node InfiniBand/RoCE fabrics, \project models service time in $O(1)$ closed form:
\begin{equation}
T_{\mathrm{comm}}(M)=T_{\mathrm{sync}}+
\sum_{l \in \mathcal{L}}\left(S_l\alpha_l+\frac{M_l}{\widetilde\beta_l\eta(M_l)}\right),
\label{eq:comm}
\end{equation}
where $S_l, M_l, \alpha_l$ are steps, bytes, and startup latency on level $l \in \mathcal{L}$, $\widetilde{\beta}_l = \beta_l k_l c_l / o_l$ scales link bandwidth $\beta_l$ across rails $k_l$, efficiency $c_l$, and per-node NIC oversubscription $o_l$, and $\eta(M) = M/(M+M_{1/2})$ models small-message ramp-up.
Between sparse anchors $(M_a, t_a)$ and $(M_b, t_b)$, \project interpolates linearly in physical service-time coordinates using weight $w = (T_{\mathrm{comm}}(M)-T_{\mathrm{comm}}(M_a))/(T_{\mathrm{comm}}(M_b)-T_{\mathrm{comm}}(M_a))$.

\subsection{Disaggregated Execution and Hardware Co-Design}
\label{subsec:method_disagg}

\begin{figure}[t]
\centering
\includegraphics[width=\columnwidth]{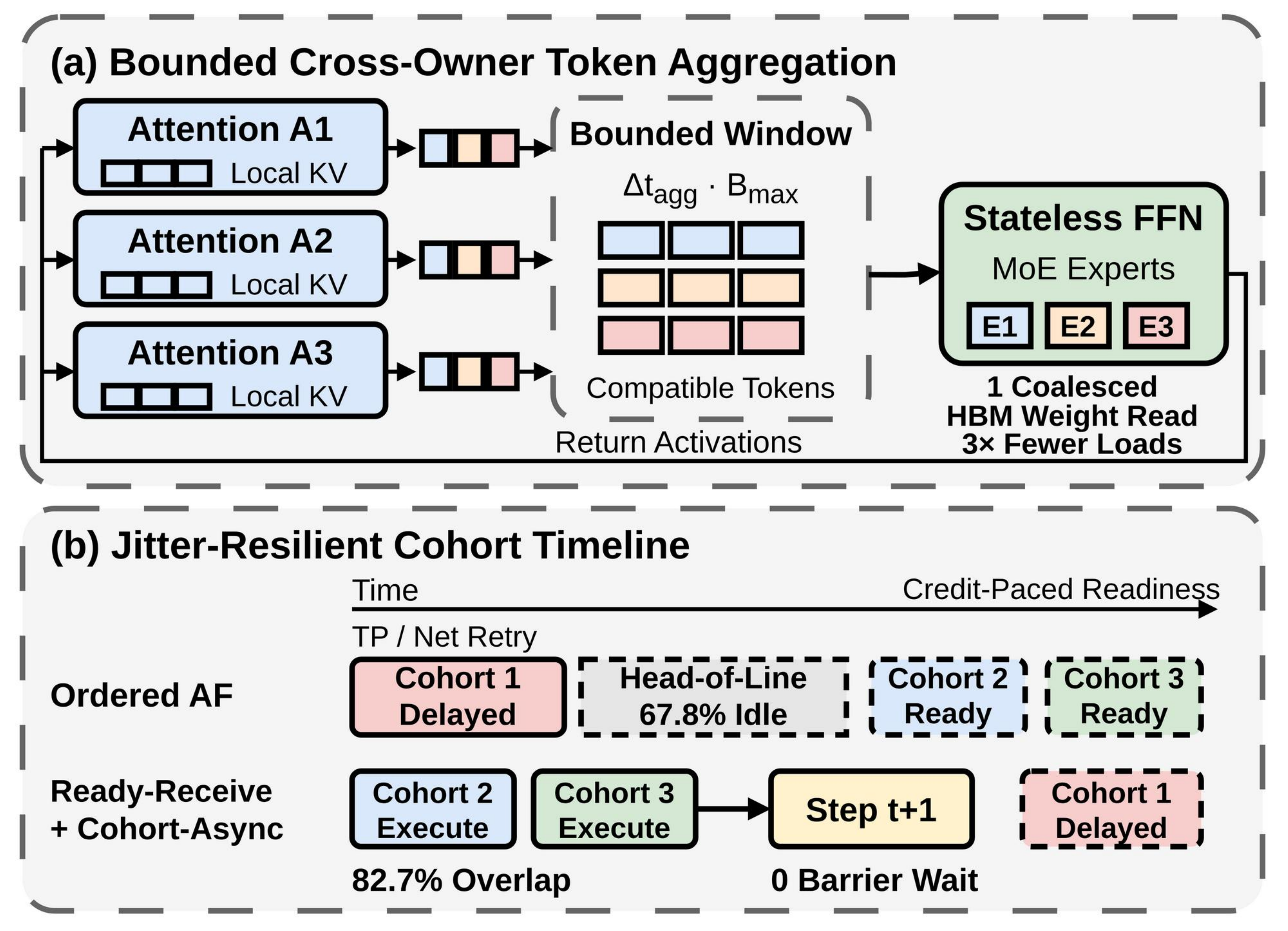}
\caption{AF execution and progress control: (a) bounded cross-owner token aggregation and (b) jitter-resilient cohort timeline.}
\label{fig:af_execution}
\vspace{-10pt}
\end{figure}

\begin{figure*}[t]
\centering
\includegraphics[width=\textwidth]{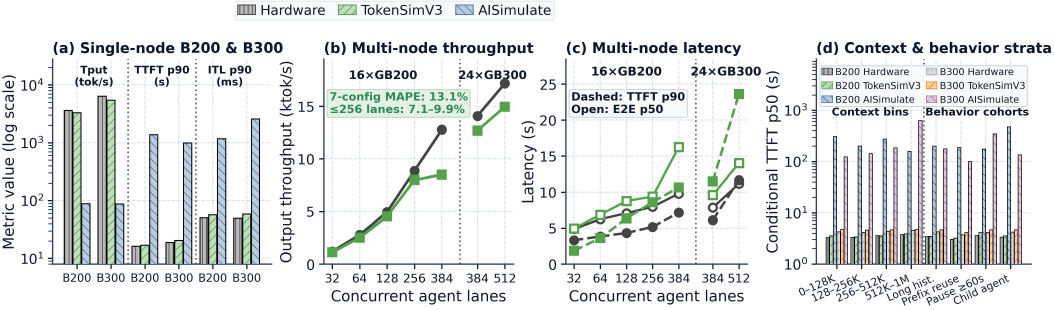}
\vspace{-20pt}
\caption{
End-to-end DeepSeek-V4 AgentX validation: (a) single-node throughput, TTFT p90, and ITL p90 on $8\times\text{B200}$ and $8\times\text{B300}$ (log scale), (b,c) multi-node P/D throughput and latency scaling, and (d) stratified conditional TTFT p50 across context bins and behavior cohorts (log scale).
}
\label{fig:agentx_accuracy}
\vspace{-10pt}
\end{figure*}

\noindent\textbf{P/D and jitter-resilient AF disaggregation.}
In P/D disaggregation~\cite{distserve,splitwise}, a completed prefill turn pins its generated \KVC{} blocks in L1 on the $P$ worker and enqueues an RDMA/NVLink transfer to the target $D$ worker or shared L3 store; the request enters the $D$ decode batch only after destination slots are reserved and the transfer finishes across contended links, capturing how compute ($r_C$) and bandwidth ($r_B$) asymmetry govern pool sizing.
AF disaggregation~\cite{step3,megascale-infer} assigns stateful attention to $A$ workers and stateless routed experts to $F$ workers, but naive scheduling stalls at high concurrency.
TP jitter and network tails reorder cohort arrivals across successive hops, turning overlap into single-batch blocking at $\ge 256$ concurrent requests and capping replicas at $32$--$64$ concurrency.
Guided by these findings, we introduce three optimizations in \project (Figure~\ref{fig:af_execution}): (1) \emph{short-window FFN batching} $(\Delta t_{\mathrm{agg}}, B_{\max})$ on $F$ workers (Figure~\ref{fig:af_execution}(a)) to merge token micro-batches across $A$ workers ($B_{\mathrm{agg}} \le B_{\max}$); (2) \emph{ready-first dispatch} (Figure~\ref{fig:af_execution}(b)) to execute arriving cohorts in readiness order rather than send order; and (3) \emph{asynchronous step continuation} (Figure~\ref{fig:af_execution}(b)) so completed sub-batches start the next decode step immediately without waiting at a full-batch barrier.

\noindent\textbf{Emerging hardware onboarding (LPX and HBF).}
Declarative hardware profiles and storage plugins extend \project without changing its runtime.
To integrate SRAM-centric LPX accelerators such as Groq LP30~\cite{groq-tsp,groq3_lpx_architecture}, developers provide a declarative hardware profile specifying per-chip compute, on-chip SRAM capacity and bandwidth, and inter-chip C2C link parameters; because LPX chips keep all state on chip without off-chip HBM, \project enforces a strict on-chip SRAM residency check to screen candidate $(\text{TP}, \text{PP}, B)$ placements before execution and estimates operator and inter-chip communication latency via the analytical \RoofModel and collective equations, calibrating a single operator-boundary launch overhead $\tau$ at short context.
To integrate High-Bandwidth Flash (HBF)~\cite{hbf}, a modular L3 storage plugin within the Mooncake-style cache state machine interpolates asymmetric read/write bandwidth from $4\,\text{KiB}$ random pages to $10\,\text{MiB}$ sequential blocks under single-lane non-preemptive flash queuing, modeling both standalone L3 flash pools and HBF-packaged GPUs with reduced near-HBM capacity.
Because smaller near-HBM triggers frequent L1 evictions and non-preemptive flash writes stall foreground restores under naive FIFO queuing, we introduce three HBF I/O optimizations in \project:
(1) \emph{selective flash write filtering} admits blocks after repeated hits ($\ge 2$) or during active parent tool pauses, dropping one-time suffixes;
(2) \emph{read-first scheduling and write rate limiting} prioritizes restores over pending writes and caps background write rate;
and (3) \emph{prefix prefetching during tool pauses}, which uses observed pause times to load a paused parent's \KVC{} from HBF into near-HBM within a memory budget before resume.
\section{Evaluation}
\label{sec:eval}

\noindent\textbf{Experimental setup.}
We evaluate \project across five models and seven platforms:
\textbf{DeepSeek-V4-Pro}~\cite{deepseekv4} (compressed + SWA attention, mixed FP8/MXFP4, EP8/EP16, MTP) on single-node $8\times\text{B200}$/$8\times\text{B300}$, multi-node P/D $16\times\text{GB200}$ ($8\text{P}:8\text{D}$)/$24\times\text{GB300}$ ($8\text{P}:16\text{D}$), and $16\times\text{B200}$ AF clusters;
\textbf{Qwen3-30B-A3B}~\cite{qwen3} on $8$--$12\times\text{B200}$ for P/D, AF, and HBF co-design;
\textbf{Qwen3-235B-A22B-FP8}~\cite{sglang_million_context} on $32\times\text{H20}$ ($\text{PP8}\times\text{TP4}$) for $1\text{M}$-context prefill;
\textbf{Gemma-4-31B}~\cite{gemma4} on a 256-chip Groq LP30 rack ($\text{TP16}\times\text{PP16}$)~\cite{groq3_lpx_architecture,artificial_analysis};
and \textbf{Llama-3.1-70B} on $4\times\text{H100}$ for simulator speed.
Workloads include the production AgentX trace ($393$ agent trees, $98,827$ dependent turns with tool pauses and sub-agent joins) and high-fan-out research trees ($262\text{K}$ parent prefixes with $4$--$8$ children appending $32\text{K}$--$786\text{K}$ tokens).
We compare against native Vidur~\cite{vidur}, LLMServingSim 2.0~\cite{llmservingsim2}, AISimulate/AIConfigurator~\cite{aisimulate,aiconfigurator}, SimAI~\cite{simai}, and ASTRA-sim 2.0~\cite{astrasim} on one pinned CPU core.

\subsection{End-to-End Cluster Fidelity and Simulator Efficiency}
\label{subsec:eval_e2e}

\begin{figure}[t]
\centering
\includegraphics[width=\columnwidth]{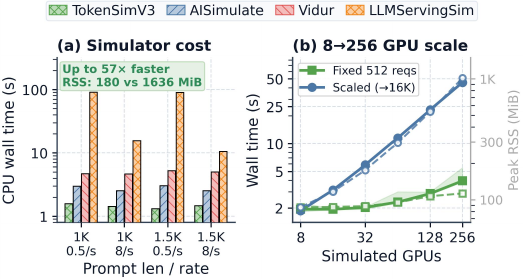}
\vspace{-20pt}
\caption{Simulator efficiency: (a) single-core wall time and peak memory on Llama-3.1-70B, and (b) scaling to $256$ GPUs.}
\label{fig:simulation_cost}
\vspace{-10pt}
\end{figure}

\begin{figure*}[t]
\centering
\includegraphics[width=\textwidth]{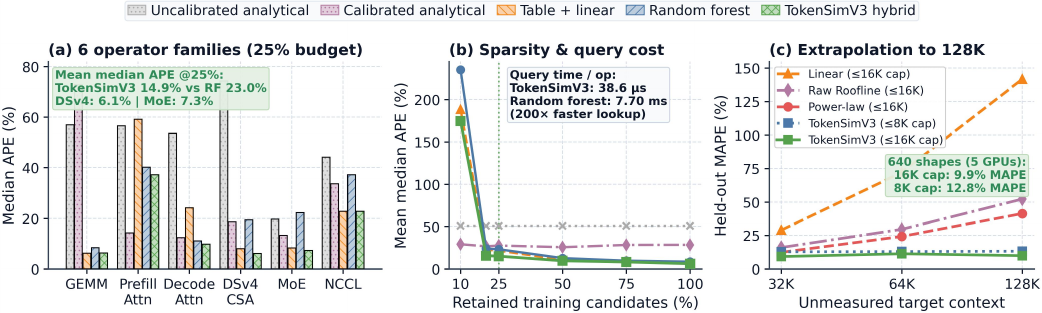}
\vspace{-20pt}
\caption{Operator latency evaluation: (a) median error across $6$ operator families at $25\%$ budget, (b) accuracy vs.\ sampling fraction and query cost, and (c) long-context decode-attention extrapolation to $128\text{K}$.}
\label{fig:operator_comparison}
\vspace{-10pt}
\end{figure*}

\noindent\textbf{Single-node, multi-node, and stratified accuracy.}
End-to-end validation covers aggregate cluster performance and latency differences across context lengths and agent behaviors.
Figure~\ref{fig:agentx_accuracy} covers single-node B200/B300 and multi-node P/D GB200 ($32$--$384$ lanes)/GB300 configurations with $140\,\text{GiB}$ host DRAM/rank.
Because Vidur~\cite{vidur} and LLMServingSim 2.0~\cite{llmservingsim2} do not support DeepSeek-V4's heterogeneous attention or global EP, Figure~\ref{fig:agentx_accuracy}(a) compares \project against AISimulate and physical hardware on single-node $8\times\text{B200}$ and $8\times\text{B300}$.
\project matches hardware across output throughput ($3,302$ vs.\ $3,599\,\text{tok/s}$ and $5,408$ vs.\ $6,360\,\text{tok/s}$, within $8.26\%$ and $14.97\%$), TTFT p90 ($16.98$ vs.\ $16.21\,\text{s}$ and $20.52$ vs.\ $18.92\,\text{s}$, within $4.78\%$ and $8.49\%$), and ITL p90 ($56.61$ vs.\ $50.25\,\text{ms}$ and $58.14$ vs.\ $49.49\,\text{ms}$), whereas native AISimulate underpredicts throughput by $97.6\%$--$98.6\%$ ($88.0/87.1\,\text{tok/s}$) and overpredicts TTFT p90 ($1,383/988\,\text{s}$) and ITL p90 ($1,169/2,562\,\text{ms}$) by $23\times$--$85\times$ due to lacking host DRAM offload, mixed FP8/MXFP4 kernels, and MTP speculation.
Across $188,580$ multi-node requests, \project tracks throughput within \textbf{$7.1\%$--$9.9\%$ error} at $32$--$256$ GB200 lanes and $384$ GB300 lanes ($13.1\%$ seven-configuration MAPE), following TTFT p90 and end-to-end latency scaling (Figure~\ref{fig:agentx_accuracy}(b,c)).
In Figure~\ref{fig:agentx_accuracy}(b,c), hardware throughput scales nearly linearly up to $256$ lanes on $16\times\text{GB200}$ ($8\text{P}:8\text{D}$) before bending at $384$ lanes as decode batching and host-DRAM offload queues saturate, whereas expanding to $24\times\text{GB300}$ ($8\text{P}:16\text{D}$) doubles decode capacity and restores scaling through $512$ lanes; \project reproduces both the linear regime and the saturation knee.
Because aggregate cluster averages can mask long-context tail errors and prefix-reuse dynamics, Figure~\ref{fig:agentx_accuracy}(d) stratifies requests across four context bins up to $1\text{M}$ tokens and four agent cohorts.
Across these strata, \project predicts conditional TTFT p50 within \textbf{$0.20\%$--$7.47\%$} on B200 and \textbf{$2.77\%$--$11.44\%$} on B300 (vs.\ $2,577\%$--$13,815\%$ for AISimulate) and tracks the TTFT p90 drop from $16.09\text{s}/16.96\text{s}$ to $6.40\text{s}/7.19\text{s}$ ($6.83\text{s}/7.93\text{s}$ simulated) on cached $64\text{K}$--$128\text{K}$ turns.
Hardware TTFT p50 stays near $3$--$5\,\text{s}$ even in the $512\text{K}$--$1\text{M}$ bin and across tool-pause and child-agent cohorts because high prefix reuse turns long histories into short incremental prefills; without host-DRAM offloading and prefix-aware cached prefill timing, AISimulate forces full recomputation and admission-queue buildup, pushing TTFT above $100\,\text{s}$ in every stratum.
These results confirm that coupling stateful agent replay, host-DRAM prefix offloading, and role-accurate kernels is essential to predict both cluster saturation points and per-cohort latency distributions.

\noindent\textbf{Simulation speed and 256-GPU scalability.}
To assess sweep cost, we measure simulation time and memory as cluster size and workload grow.
As shown in Figure~\ref{fig:simulation_cost}(a), on Llama-3.1-70B ($4\times\text{H100}$ TP4), \project finishes in $1.32$--$1.57\,\text{s}$, achieving $1.9\times$--$2.0\times$ speedup over AISimulate ($2.52$--$3.03\,\text{s}$), $3.3\times$--$3.5\times$ over Vidur ($4.65$--$5.24\,\text{s}$), and $7.9\times$--$57.3\times$ over LLMServingSim 2.0 ($10.44$--$89.90\,\text{s}$), with $9.0\times$ lower peak RSS than Vidur ($182\,\text{MiB}$ vs.\ $1637\,\text{MiB}$).
Figure~\ref{fig:simulation_cost}(b) shows that when scaling Qwen3-30B-A3B ($\text{TP4}$ replicas) from $8$ to $256$ GPUs on one core, strong-scaling wall time ($512$ fixed requests) grows sub-linearly from $1.94\,\text{s}$ to $3.94\,\text{s}$, and weak scaling ($64$ requests/GPU, up to $16,384$ requests at $256$ GPUs) completes in $45.61\,\text{s}$ within $1.01\,\text{GiB}$ RSS.
Cycle-level stepping in LLMServingSim 2.0 inflates runtime especially at lower arrival rates ($0.5/\text{s}$) where simulated wall-clock spans lengthen, while Vidur's Random Forest ensembles drive $9.0\times$ higher process memory; by pairing event-driven execution with $O(1)$ operator lookups, \project keeps single-core sweeps under a minute even at $256$ GPUs.

\subsection{Operator and Collective Engine Validation}
\label{subsec:eval_operator}

\begin{figure*}[t]
\centering
\begin{minipage}[c]{0.49\textwidth}
\centering
\includegraphics[width=\linewidth]{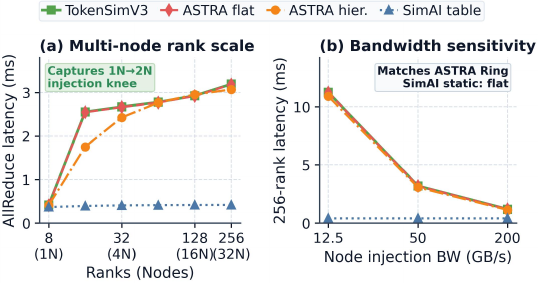}
\end{minipage}\hfill
\begin{minipage}[c]{0.49\textwidth}
\centering
\includegraphics[width=\linewidth]{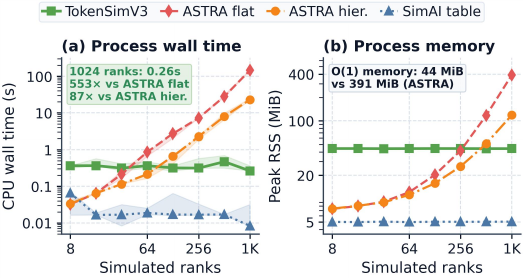}
\end{minipage}
\vspace{-10pt}
\caption{Multi-node collective topology fidelity and simulator scalability: \textbf{Left:} rank scaling ($8$--$256$ ranks) capturing the $1\text{N}\to 2\text{N}$ NIC injection knee and node injection bandwidth sensitivity ($12.5$--$200\,\text{GB/s}$). \textbf{Right:} single-core CPU wall time and peak memory for $64$ sequential $1\,\text{MiB}$ AllReduce calls from $8$ to $1024$ ranks ($553\times$ speedup over ASTRA-sim flat Ring).}
\label{fig:comm_scale}
\vspace{-10pt}
\end{figure*}

\noindent\textbf{Sparse $25\%$ profiling and $1\text{M}$-context extrapolation.}
The operator study tests sparse profiling and physical scaling on unmeasured shapes, cached prefill, and long contexts.
Figure~\ref{fig:operator_comparison}(a,b) evaluates operator latency prediction across $12$ B200/B300 operator strata ($6$ operator families: GEMM, prefill/decode attention, compressed attention, MoE, and NCCL collectives) across $10\%$--$100\%$ sampling fractions: at a $25\%$ budget, \project achieves the lowest mean median error (\textbf{$14.87\%$}), outperforming linear table extrapolation ($21.41\%$), Vidur Random Forest ($23.04\%$), calibrated Roofline ($27.46\%$), and uncalibrated Roofline ($50.83\%$).
As shown in Figure~\ref{fig:operator_comparison}(a,b), on non-linear kernels, \project cuts B200/B300 median error to $6.3\%/5.8\%$ on compressed attention and $6.9\%/7.6\%$ on MoE layers while executing in $38.6\,\mu\text{s}$ per query, \textbf{$200\times$ faster} than Random Forest ($7.70\,\text{ms}$).
Two trends explain this advantage: in Figure~\ref{fig:operator_comparison}(a), pure rooflines miss kernel-specific wave quantization and MoE dispatch inefficiencies ($20\%$--$70\%$ error), whereas sparse local anchors capture realized kernel behavior directly; in Figure~\ref{fig:operator_comparison}(b), error drops sharply once sampling reaches a $25\%$ knee and plateaus thereafter, cutting profiling effort by $4\times$ with minimal accuracy loss.
Figure~\ref{fig:operator_comparison}(c) shows that when extrapolating $8\times$--$16\times$ to $32\text{K}$--$128\text{K}$ decode-attention across five GPU architectures from sparse single-node $\le 16\text{K}$ or $\le 8\text{K}$ profiles, isolating fixed launch overhead (Equations~\ref{eq:extrap}--\ref{eq:roofline}) yields a flat \textbf{$9.92\%$ and $12.83\%$ MAPE}, whereas linear extrapolation scales short-context launch overhead with sequence length and diverges to $65.28\%$--$144.74\%$, and raw roofline or power-law fits drift to $40\%$--$52\%$.
Similarly, as shown in Figure~\ref{fig:estimator_flow} (right), historical-context correction halves off-diagonal cached-prefill MAPE to $8.0\%$--$8.7\%$ using only diagonal anchors, and anchoring Qwen3-235B-A22B-FP8 chunked prefill ($\text{PP8}\times\text{TP4}$ on $32\times\text{H20}$~\cite{sglang_million_context}) at a single $128\text{K}$ point predicts held-out $256\text{K}$, $512\text{K}$, and $1\text{M}$ TTFT ($32.68\,\text{s}$, $114.33\,\text{s}$, $420.91\,\text{s}$) within \textbf{$0.40\%$ MAPE} vs.\ $104.2\%$ for raw roofline.
Together, sparse local anchoring and physics-guided scaling eliminate the accuracy--cost--extrapolation trade-off across million-token contexts and cached prefixes.

\noindent\textbf{Multi-node collective accuracy and $553\times$ speedup.}
To verify whether closed-form hierarchical modeling preserves cluster-scale fabric bottlenecks without incurring packet-level overhead, we evaluate collective prediction across message sizes, node counts, and NIC injection bandwidths.
Table~\ref{tab:comm_error} evaluates collective latency accuracy on held-out message sizes ($\ge 4\,\text{KiB}$) across $83$ vendor-published NCCL points using $1$-in-$4$ sparse anchors, where stock SimAI and ASTRA-sim 2.0 incur $59.65\%$--$92.63\%$ and $26.07\%$--$64.37\%$ MAPE because fixed formulas miss NCCL protocol transitions (LL, LL128, Simple) and small-message ramp-up.
Interpolating in physical service-time coordinates (Equation~\ref{eq:comm}) reduces held-out MAPE to \textbf{$9.27\%$ on AWS, $5.23\%$ on Google, and $7.39\%$ on NVIDIA}, cutting error by $52.4\%$--$85.7\%$ compared to axis-coordinate interpolation over the same anchors.
Across $144$ topology configurations spanning $8$--$256$ ranks ($1\text{N}$--$32\text{N}$) and $12.5$--$200\,\text{GB/s}$ per-node NIC injection bandwidth (Figure~\ref{fig:comm_scale}, left), \project reproduces the sharp $1\text{N}\to 2\text{N}$ NIC injection knee ($0.4\to 2.6\,\text{ms}$) and nearly $10\times$ bandwidth scaling ($11\to 1.1\,\text{ms}$) of ASTRA-sim where static SimAI tables remain completely flat because they ignore inter-node NIC oversubscription.
Meanwhile, Figure~\ref{fig:comm_scale} (right) shows that packet-level simulation in ASTRA-sim grows super-linearly with rank count, whereas evaluating Equation~\ref{eq:comm} in closed form completes $64$ sequential AllReduce calls at $1024$ ranks in \textbf{$0.264\,\text{s}$ with constant $O(1)$ peak memory ($44.1\,\text{MiB}$ vs.\ $391\,\text{MiB}$ for ASTRA-sim)}, achieving a \textbf{$553\times$ speedup} over ASTRA-sim flat Ring ($146.10\,\text{s}$) and $86.7\times$ over hierarchical Ring ($22.90\,\text{s}$).
Thus, physical-coordinate anchoring combined with closed-form hierarchical contention modeling delivers multi-node topology fidelity without packet-level simulation cost.

\begin{table}[t]
\centering
\caption{Held-out multi-node collective latency MAPE (\%) across vendor-published NCCL benchmarks.}
\label{tab:comm_error}
\papertablestyle
\renewcommand{\arraystretch}{1.05}
\begin{tabularx}{\columnwidth}{@{}L{1}rrr@{}}
\toprule
Predictor & AWS$^\dagger$ & Google$^\ddagger$ & NVIDIA$^\S$ \\
\midrule
SimAI (stock table)~\cite{simai} & 92.63 & 59.65 & 68.84 \\
ASTRA-sim 2.0 (flat Ring)~\cite{astrasim} & 64.26 & 47.88 & 26.07 \\
ASTRA-sim 2.0 (hier.\ Ring)~\cite{astrasim} & 64.37 & 55.50 & 47.90 \\
\project (axis-coord.\ anchors) & 19.47 & 36.48 & 34.69 \\
\textbf{\project (physical anchors)} & \textbf{9.27} & \textbf{5.23} & \textbf{7.39} \\
\bottomrule
\multicolumn{4}{@{}p{\columnwidth}@{}}{\scriptsize $^\dagger$AWS $2\times\text{P4d}$ ($16\times\text{A100}$ EFA)~\cite{aws_nccl_efa}; $^\ddagger$Google Cloud $2\times\text{A4}$ ($16\times\text{B200}$ gIB)~\cite{google_nccl_vms}; $^\S$NVIDIA BasePOD $4\times\text{DGX}$ ($32\times\text{H100}$ IB)~\cite{nvidia_basepod_nccl}.}
\end{tabularx}
\vspace{-12pt}
\end{table}

\subsection{Programmable \KVC{} Exploration}
\label{subsec:eval_kvcache}

\begin{figure}[t]
\centering
\includegraphics[width=\columnwidth]{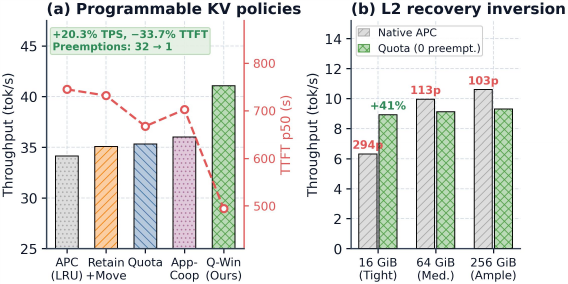}
\vspace{-14pt}
\caption{Programmable \KVC{} exploration on $8\times\text{B200}$ (DeepSeek-V4-Pro): (a) programmable retention, movement, and quota primitives over passive LRU (\texttt{APC}) and (b) L2 host DRAM recovery capacity inversion.}
\label{fig:pkv_policies}
\vspace{-10pt}
\end{figure}

Production serving engines rely on fixed LRU eviction and static offload heuristics, whereas programmable \KVC{} interfaces expose turn-boundary hooks for custom retention, movement, quota, and write-back logic (Section~\ref{subsec:method_pkv}).
We use \project to explore this programmable \KVC{} design space on $8\times\text{B200}$ running DeepSeek-V4-Pro ($\text{EP8}$, MTP enabled) without modifying a live inference stack, comparing passive LRU (\texttt{APC}) against composable programmable primitives under tool pauses and multi-branch contention.

\noindent\textbf{Programmable retention, movement, and quota exploration.}
Static tool-wait timers ignore cluster prefill queuing and can evict a paused parent prefix just before a queued turn resumes.
Figure~\ref{fig:pkv_policies}(a) explores progressive programmable cache primitives across $64$ fan-out agent trees ($262\text{K}$ shared parent prefix, four child branches appending $131\text{K}$ private tokens, $120\,\text{s}$ tool pauses), where passive \texttt{APC} suffers $32$ preemptions, $4.33\text{M}$ recomputed tokens, $745.1\,\text{s}$ TTFT p50, and $34.14\,\text{tok/s}$.
Composing fixed retention-and-movement (\texttt{Retain+Move}), per-tree \texttt{Quota}, and lifecycle hints (\texttt{App-Coop}) yields modest gains ($35.08$--$35.99\,\text{tok/s}$) because static timers still hold stale prefixes too long or demote parents during prefill queuing.
Programming the retention hook with \project's queue-coupled window $\widehat{T}_{\mathrm{window}}$ (\texttt{Q-Win}) cuts preemptions from $32$ to $1$, raises throughput to \textbf{$41.07\,\text{tok/s}$ ($+20.3\%$)}, and lowers TTFT p50 by \textbf{$33.7\%$ to $494.2\,\text{s}$} (reaching $466.7\,\text{s}$, $-37.4\%$, when paired with post-join child GPU reclamation at $39.98\,\text{tok/s}$).
On $196$-lane AgentX replay, programmable queue-coupled retention sustains $1,970$--$1,993\,\text{tok/s}$ ($71.5\%$--$80.9\%$ GPU hit rate), whereas unconditional $30\,\text{s}$ forced demotion drops GPU hit rate to $20.2\%$, triggers $388.8\,\text{GiB}$ of host-to-GPU reloads, and lowers throughput to $1,941.3\,\text{tok/s}$.
Exploring programmable quota admission across $16, 64,$ and $256\,\text{GiB/rank}$ of L2 host DRAM on $32$ deep-research trees (eight $786\text{K}$-token children per tree; Figure~\ref{fig:pkv_policies}(b)) exposes a capacity-dependent inversion.
Under tight L2 ($16\,\text{GiB/rank}$), concurrent child branches overwrite evicted parent prefixes under passive \texttt{APC}, causing $294$ preemptions ($156.5\text{M}$ recomputed tokens); programmable per-tree quotas eliminate preemptions, lifting throughput by \textbf{$+41.1\%$} ($8.91$ vs.\ $6.31\,\text{tok/s}$) and cutting TTFT p50 by \textbf{$40.5\%$}.
Under ample L2 ($64$--$256\,\text{GiB/rank}$), fast PCIe/C2C restores amortize GPU preemptions, so optimistic \texttt{APC} ($9.96$--$10.60\,\text{tok/s}$) outperforms conservative static quotas ($9.11$--$9.30\,\text{tok/s}$) that restrict concurrency.

\begin{figure}[t]
\centering
\includegraphics[width=\columnwidth]{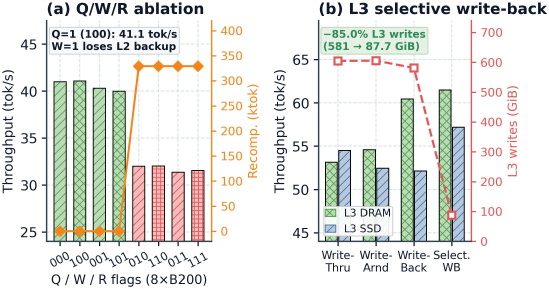}
\vspace{-14pt}
\caption{Programmable \KVC{} tier-boundary exploration on $8\times\text{B200}$ (DeepSeek-V4-Pro): (a) programmable \texttt{Q/W/R} hook ablation and (b) programmable L2$\to$L3 write-back policies across DRAM and SSD tiers.}
\label{fig:pkv_tiering}
\vspace{-10pt}
\end{figure}

\noindent\textbf{Programmable L1/L2/L3 write-back boundary exploration.}
Completed child suffixes are rarely reused across turns, yet programming a write bypass at the wrong cache boundary risks losing active in-flight work upon scheduler preemption.
In the eight-configuration hook ablation of Figure~\ref{fig:pkv_tiering}(a), enabling the L1$\to$L2 write-bypass hook (\texttt{W=1}) creates a performance cliff across every combination: preempted child branches lose their $33\text{K}$ private tokens, inflating recomputed prefill tokens by $1,296\times$ ($254 \to 329,214$) and dropping throughput by $22.1\%$ ($41.07 \to 31.99\,\text{tok/s}$).
Figure~\ref{fig:pkv_tiering}(b) instead evaluates programmable write-back hooks at the L2$\to$L3 boundary: write-through and write-around saturate shared L3 bandwidth with $\sim 605\,\text{GiB}$ of traffic, and standard write-back improves L3 DRAM ($60.4\,\text{tok/s}$) while lagging on slower L3 SSD ($52.1\,\text{tok/s}$).
Preserving L2 recovery backups for active branches while programming selective one-shot filtering strictly at L2$\to$L3 eviction raises L3 DRAM throughput to \textbf{$61.49\,\text{tok/s}$ ($+15.7\%$ over write-through)}, raises L3 SSD throughput to $57.18\,\text{tok/s}$, and \textbf{cuts L3 write volume by $85.0\%$} ($581.4 \to 87.7\,\text{GiB}$) without hurting L3 prefix restores.
This programmable \KVC{} exploration establishes three rules for cache-controller APIs: retention hooks must track live prefill queues, quota hooks should activate under host-DRAM pressure, and one-shot write filters belong strictly at the L2$\to$L3 boundary.

\subsection{Disaggregated P/D and AF Serving Co-Design}
\label{subsec:eval_disagg}

\begin{figure}[t]
\centering
\includegraphics[width=\columnwidth]{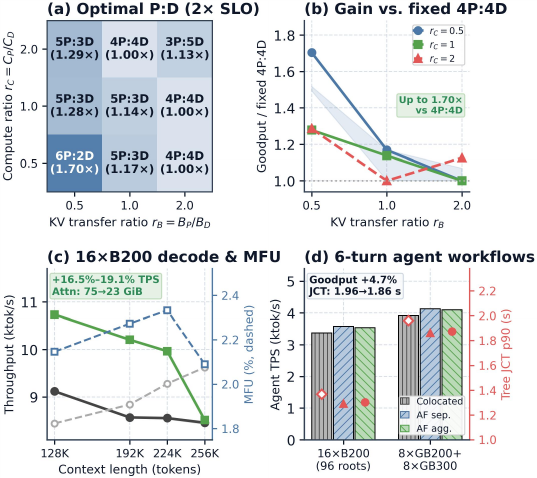}
\vspace{-16pt}
\caption{
P/D and AF disaggregation: (a) optimal P:D split and (b) goodput gain on Qwen3-30B-A3B; (c) decode speed/MFU and (d) $6$-turn agent goodput/JCT p90 on DeepSeek-V4-Pro.
}
\label{fig:af_benefit}
\vspace{-14pt}
\end{figure}

\noindent\textbf{P/D ratio shift and AF memory-capacity gain.}
To test whether fixed pool ratios remain optimal under asymmetric hardware and how expert offloading affects long-context MoE serving, we sweep P/D worker splits and AF placements under identical GPU budgets.
Figure~\ref{fig:af_benefit}(a,b) sweeps $819$ P/D splits on $8$ Qwen3-30B-A3B workers across compute ($r_C \in \{0.5, 1.0, 2.0\}$) and bandwidth ($r_B \in \{0.5, 1.0, 2.0\}$) ratios: fixed $4\text{P}:4\text{D}$ is optimal in only $3$ of $9$ regimes, because constrained prefill compute and transfer links ($r_C = r_B = 0.5$) make prefill queuing and \KVC{} handoff dominate TTFT ($6\text{P}:2\text{D}$ yields \textbf{$1.70\times$ higher SLO goodput}), constrained transfer links alone ($r_C = 1.0, r_B = 0.5$) still require $5\text{P}:3\text{D}$ ($1.28\times$ gain) as KV egress stalls tie up prefill ranks, and faster prefill hardware ($r_C = r_B = 2.0$) shifts the bottleneck to decode concurrency ($3\text{P}:5\text{D}$, $1.18\times$ gain).
Figure~\ref{fig:af_benefit}(c,d) evaluates MoE AF disaggregation on $61$-layer DeepSeek-V4-Pro across $16$ GPUs: a $2\text{A}:14\text{F}$ split improves decode throughput by \textbf{$16.5\%$--$19.1\%$} at $128\text{K}$--$224\text{K}$ context ($8,569 \to 10,204\,\text{tok/s}$ at $192\text{K}$ on $16\times\text{B200}$, Figure~\ref{fig:af_benefit}(c)) while lifting multi-turn agent goodput by \textbf{$+4.66\%$} and cutting JCT p90 by \textbf{$5.05\%$} ($1.96 \to 1.86\,\text{s}$) on $8\times\text{GB200}+8\times\text{GB300}$ (Figure~\ref{fig:af_benefit}(d)), as offloading routed experts shrinks $A$-rank weights from $74.56$ to $22.96\,\text{GiB}$, doubling attention block size ($64 \to 128$) and halving FFN calls ($62,464 \to 27,328$).
Beyond $224\text{K}$ context ($256\text{K}$ in Figure~\ref{fig:af_benefit}(c)), AF and colocated curves converge once attention KV bandwidth saturates even the freed $A$-rank HBM, delineating the exact context window where $2\text{A}:14\text{F}$ pays off.

\begin{figure}[t]
\centering
\includegraphics[width=\columnwidth]{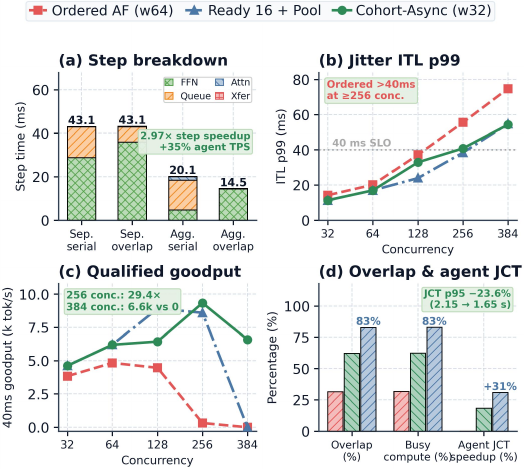}
\vspace{-16pt}
\caption{
AF aggregation and jitter resilience on Qwen3-30B-A3B: (a) step latency breakdown, (b) ITL p99, (c) $40\,\text{ms}$-qualified goodput, and (d) overlap and agent JCT p95 at $256$ concurrency.
}
\label{fig:ffn_aggregation}
\vspace{-14pt}
\end{figure}

\noindent\textbf{FFN token batching and high-concurrency jitter resilience.}
Because AF disaggregation splits each layer across remote pools, we examine both small-batch FFN weight-streaming overhead and high-concurrency ordering stalls under straggler jitter.
As shown in Figure~\ref{fig:ffn_aggregation}(a), on Qwen3-30B-A3B across $8\times\text{B200}$ ($6\text{A}:2\text{F}$), pipeline overlap alone yields zero speedup ($43.1\,\text{ms}$ serial vs.\ overlap) because uncoalesced FFN HBM weight streaming ($36\,\text{ms}$) dominates the step; merging micro-batches across a $5\,\mu\text{s}$ window on $F$ workers cuts expert-weight reads by \textbf{$3.0\times$} ($192 \to 64$ loads), shrinking FFN time so overlap hides queuing and cuts step latency by \textbf{$2.97\times$} ($43.09 \to 14.50\,\text{ms}$, \textbf{$+35.3\%$ agent throughput}).
Figure~\ref{fig:ffn_aggregation}(b--d) tests $32$--$384$ concurrent requests on $12\times\text{B200}$ ($4\text{P}+4\text{A}+4\text{F}$) under TP jitter ($\text{CV}=0.25$) and $1\%$ fabric retries.
Out-of-order $A \leftrightarrow F$ arrivals cause single-batch blocking (Section~\ref{subsec:method_disagg}), capping replicas at $32$--$64$ concurrency and leaving pipelines idle for $67.8\%$ of wall time at $256$.
The $40\,\text{ms}$-qualified goodput drops from $4,828\,\text{tok/s}$ ($64$) to $317\,\text{tok/s}$ ($256$) and $0\,\text{tok/s}$ ($384$).
Ready-first dispatch and $5\,\mu\text{s}$ FFN batching eliminate intra-step ordering stalls—cutting $256$-concurrency idle time to $23.4\%$, lifting overlap from $31.5\%$ to \textbf{$82.7\%$}, restoring qualified goodput by \textbf{$27.1\times$} ($8,596\,\text{tok/s}$), and cutting agent JCT p95 by \textbf{$23.6\%$}—yet still collapse at $384$ concurrency due to end-of-step owner barriers.
Adding asynchronous step continuation removes that inter-step barrier, reaching \textbf{$9,334\,\text{tok/s}$ ($29.4\times$)} at $256$ and sustaining \textbf{$6,579\,\text{tok/s}$} at $384$ concurrency.
These results demonstrate that disaggregated serving realizes its capacity and MFU gains only when pool splits track $(r_C, r_B)$ asymmetry and AF pipelines pair cross-owner FFN batching with ready-first dispatch and asynchronous step continuation.

\subsection{Emerging Hardware Exploration: LPX and HBF}
\label{subsec:eval_frontier}

\begin{figure}[t]
\centering
\includegraphics[width=\columnwidth]{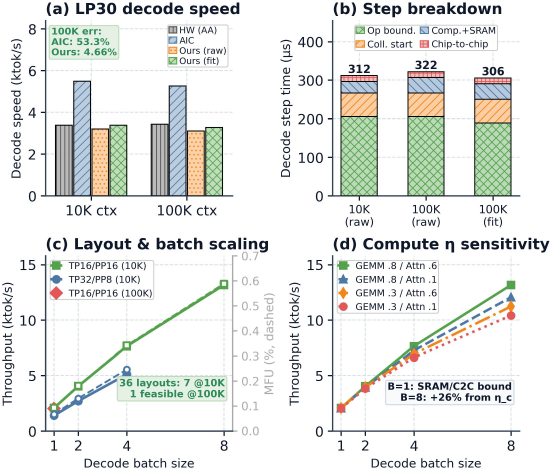}
\vspace{-16pt}
\caption{256-chip Groq LP30 on Gemma-4-31B: (a) decode speed, (b) step breakdown, (c) $\text{TP}\times\text{PP}$ and batch scaling, and (d) $\eta_c$ sensitivity.}
\label{fig:lpx_utilization}
\vspace{-14pt}
\end{figure}

\noindent\textbf{SRAM-centric LP30 accuracy and topology search.}
To validate \project on non-GPU accelerators where profiled kernel tables are unavailable, we evaluate a 256-chip SRAM-centric LP30 rack~\cite{groq3_lpx_architecture} running Gemma-4-31B~\cite{gemma4} against Artificial Analysis measurements at $10\text{K}$/$100\text{K}$ context ($3,382$/$3,431\,\text{tok/s}$)~\cite{artificial_analysis}.
Native AIConfigurator overpredicts throughput by $53.3\%$--$62.1\%$, whereas \project achieves \textbf{$5.20\%$ and $9.46\%$ error} uncalibrated and \textbf{$4.66\%$ held-out error at $100\text{K}$} after calibrating a single operator-boundary launch parameter $\tau$ ($0.200 \to 0.184\,\mu\text{s}$) at $10\text{K}$ (Figure~\ref{fig:lpx_utilization}(a)).
Decomposing the $311.8\,\mu\text{s}$ step at $B=1$ explains both why raw rooflines fail on SRAM processors and why hardware decode speed stays nearly constant from $10\text{K}$ to $100\text{K}$ ($312 \to 322\,\mu\text{s}$, $+3.2\%$): $150\,\text{TB/s}$ SRAM compresses GEMM and attention execution to $45.6\,\mu\text{s}$ ($14.6\%$), leaving per-layer operator boundary transitions ($205.2\,\mu\text{s}$, $65.8\%$) and inter-chip collective startup ($61.0\,\mu\text{s}$, $19.6\%$) as $85.4\%$ of total latency (Figure~\ref{fig:lpx_utilization}(b)).
Across $36$ candidate $(\text{TP}, \text{PP}, B)$ placements, the $500\,\text{MB/chip}$ SRAM residency check filters feasible layouts down to $7$ at $10\text{K}$ and $1$ at $100\text{K}$—rejecting $\text{TP8}\times\text{PP32}$ (weight OOM) and $\text{TP32}\times\text{PP8}$ (\KVC{} OOM)—and selects $\text{TP16}\times\text{PP16}$, which scales throughput by $6.33\times$ ($2,078 \to 13,161\,\text{tok/s}$) at $B=8$ while rack MFU remains below $0.6\%$ because SRAM capacity spreads weights across $256$ chips (Figure~\ref{fig:lpx_utilization}(c)).
Correspondingly, Figure~\ref{fig:lpx_utilization}(d) shows that raising compute efficiency $\eta_c$ yields negligible speedup at $B=1$--$2$ because SRAM and C2C startup dominate, yet unlocks $+26\%$ higher throughput at $B=8$ as arithmetic intensity grows—guiding compilers toward operator fusion at low batch and compute optimization at larger batches.

\begin{figure}[t]
\centering
\includegraphics[width=\columnwidth]{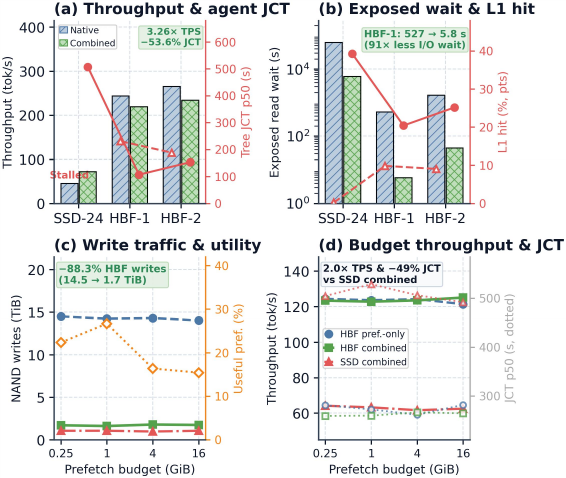}
\vspace{-16pt}
\caption{HBF tiering and I/O scheduling on $32$-turn Qwen3-30B-A3B ($8\times\text{B200}$): (a) throughput and JCT p50, (b) storage wait and L1 hit rate, and (c,d) writes, utility, and JCT vs.\ prefetch budget.}
\label{fig:hbf_policy}
\vspace{-14pt}
\end{figure}

\noindent\textbf{HBF flash tiering and workload-aware I/O scheduling.}
Replacing half of near-GPU HBM with multi-terabyte High-Bandwidth Flash (HBF) trades near-memory capacity for high far-memory bandwidth~\cite{hbf}; we evaluate when this trade-off yields net end-to-end agent gains and which I/O schedules are required to realize them.
Figure~\ref{fig:hbf_policy} compares three organizations on $8\times\text{B200}$ serving $128$ concurrent $32$-turn Qwen3-30B-A3B agents ($128\text{K}$ context): NVMe SSD-24 ($192\,\text{GB}$ HBM at $8\,\text{TB/s}$, $24\,\text{TB}$ SSD), HBF-1 ($96\,\text{GB}$ HBM at $2.69\,\text{TB/s}$, $48\,\text{TB}$ HBF at $2.4\,\text{TB/s}$), and HBF-2 ($96\,\text{GB}$ HBM at $4.0\,\text{TB/s}$, $24\,\text{TB}$ HBF at $1.2\,\text{TB/s}$).
Under native FIFO queuing, SSD-24 stalls under $>10^4\,\text{s}$ of read wait, whereas HBF boosts agent throughput by \textbf{$3.05\times$--$3.69\times$} ($219.0$--$265.3$ vs.\ $71.9\,\text{tok/s}$, Figure~\ref{fig:hbf_policy}(a)); yet halving near-HBM ($192 \to 96\,\text{GB}$) increases L1 evictions so that non-preemptive background writes still expose $527\,\text{s}$ of foreground restore wait and drive $15.93\,\text{TB}$ of NAND writes (Figure~\ref{fig:hbf_policy}(b,c)).
Enabling foreground read priority, a $16\,\text{GB/s}$ write rate limit with reuse-gated admission ($\ge 2$ hits), and bounded tool-wait prefetching cuts exposed storage wait by $91\times$ ($527 \to 5.8\,\text{s}$), reduces HBF-1 JCT p50 by \textbf{$53.6\%$} ($230.6 \to 107.0\,\text{s}$) and HBF-2 JCT p50 by $18.5\%$ ($188.0 \to 153.3\,\text{s}$), and eliminates \textbf{$88.3\%$ of NAND writes} ($15.93 \to 1.87\,\text{TB}$) with $<1\%$ throughput loss; once write blocking is removed, HBF-1 outperforms HBF-2 because its $2\times$ higher flash read bandwidth ($2.4$ vs.\ $1.2\,\text{TB/s}$) drains restore queues faster.
Sweeping the prefetch budget (Figure~\ref{fig:hbf_policy}(c,d)) shows that prefetch utility peaks near $27\%$ at $0.25$--$1\,\text{GiB}$ and declines to $\sim 16\%$ at $4$--$16\,\text{GiB}$ because oversized speculative loads are evicted before turn resumption, while combined HBF scheduling sustains $2.0\times$ higher throughput and $49\%$ lower JCT than equally scheduled SSD.
Contrasting decode-heavy coding workloads—where raising near-HBM bandwidth from $2.69$ to $8.0\,\text{TB/s}$ cuts ITL p90 from $24.53$ to $10.66\,\text{ms}$—against restoration-heavy multi-turn agents ($5.5\times$--$5.9\times$ HBF throughput gain) demonstrates how \project identifies whether an architecture should prioritize near-HBM decode bandwidth or far-memory prefix restoration.
\section{Architectural and System Insights}
\label{sec:discussion}

\subsection{System Co-Design Guidelines}
\label{subsec:disc_guidelines}

Our full-stack simulation studies across disaggregated GPU clusters and emerging hardware distill four system design guidelines for agentic LLM serving:

\noindent\textbf{Expose programmable \KVC{} policies coupled to queuing and preemption.}
Treating multi-tier offloading as an isolated storage layer governed by static timers or uniform LRU causes severe thrashing under multi-turn agent workloads (Section~\ref{subsec:eval_kvcache}).
Because an agent's true prefix reuse interval spans both external tool waits and on-cluster prefill queuing, retention policies must incorporate live queue backlogs while separating L2 preemption-recovery backups from L3 cross-turn persistence.

\noindent\textbf{Co-design disaggregated pool ratios with fabric handoff and ordering control.}
Separating P/D or AF pools eliminates colocated interference but shifts bottlenecks to inter-pool bandwidth, FFN weight-streaming fragmentation, and multi-hop ordering stalls (Section~\ref{subsec:eval_disagg}).
Effective disaggregated deployments must adapt P:D ratios to compute and interconnect asymmetry while pairing bounded cross-owner FFN batching with ready-first dispatch and asynchronous step continuation to preserve overlap under high-concurrency jitter.

\noindent\textbf{Target launch and collective startup on SRAM-centric accelerators.}
When high-bandwidth on-chip SRAM shrinks small-batch matrix and attention time to a minor share of each step, per-layer operator boundary transitions and inter-chip collective startup dominate decode latency (Section~\ref{subsec:eval_frontier}).
Hardware--software co-design for SRAM processors must therefore prioritize cross-layer kernel fusion, low-latency collective fabrics, and strict per-chip SRAM residency screening alongside batch scaling.

\noindent\textbf{Protect foreground restores when deploying far-memory flash tiers.}
High-Bandwidth Flash expands agent \KVC{} capacity to tens of terabytes, yet raw flash throughput alone fails to deliver low completion time when non-preemptive background evictions block foreground prefix reads (Section~\ref{subsec:eval_frontier}).
Realizing HBF gains requires workload-aware read prioritization, reuse-gated write admission, and bounded tool-wait prefetching tailored to whether the workload is bound by near-HBM decode bandwidth or far-memory prefix restoration.

\subsection{Accelerating Recursive Self-Improvement (RSI) Exploration}
\label{subsec:disc_rsi}

In Recursive Self-Improvement (RSI) workflows for LLM serving systems, autonomous agents continuously synthesize, test, and refine optimization algorithms and cluster deployment configurations, yet physical cluster evaluation limits iteration speed through high latency, hardware cost, and measurement variance.
With second-scale, deterministic replay on a single CPU core (Sections~\ref{subsec:eval_e2e}--\ref{subsec:eval_operator}), \project serves as a high-throughput verifier that lets RSI agents evaluate large spaces of candidate algorithms and runtime configurations without consuming production GPUs.
Through declarative cluster specifications and lightweight policy hooks, agents can inject new scheduling, routing, or caching logic with minimal code changes and receive reproducible diagnostic feedback across latency percentiles, cache hit rates, memory feasibility, and hardware utilization.
By automatically pruning infeasible or low-gain candidates in simulation and promoting only top-ranked designs to physical clusters for final validation, \project accelerates autonomous systems optimization.
\section{Conclusion}
\label{sec:conclusion}

We presented \project, a full-stack simulator for agentic and disaggregated LLM serving coupling a multi-tier \KVC{} state machine and layer-grouped allocator with a physics-anchored latency engine.
It tracks GB200/GB300 throughput within $7.1\%$--$9.9\%$ while running up to $57\times$ and $553\times$ faster than serving and packet simulators to guide \KVC{}, P/D, AF, LPX, and HBF co-design.

\bibliographystyle{ACM-Reference-Format}
\bibliography{reference}

\end{document}